\documentclass[preprint, preprintnumbers,nofootinbib,12pt]{revtex4-2}
\usepackage[breaklinks]{hyperref}
\hypersetup{colorlinks,urlcolor=black,citecolor=black,linkcolor=black,filecolor=black}
\usepackage{breakurl}
\usepackage[english]{babel}
\usepackage{color}
\usepackage{amsmath}
\usepackage{physics}
\usepackage{amssymb}
\usepackage{latexsym}
\usepackage{graphicx}
\usepackage{adjustbox}
\usepackage{xr}
\usepackage{siunitx}
\usepackage{threeparttable}
\usepackage{braket}
\usepackage[justification=centering]{caption}
\usepackage{float}
\usepackage{caption}
\usepackage{subcaption}
\usepackage{mathtools}
\usepackage{array}
\usepackage{seqsplit}
\usepackage{textcomp}
\usepackage{enumitem}
\usepackage{tikz}
\usepackage{tabularx}
\usepackage{setspace}
\usepackage{soul}
\usepackage[scr=rsfs]{mathalpha}
\usepackage{amsmath,bm}

\newcommand{\pa}{\partial}

\newcommand{\la}{\lambda}

\newcommand{\rar}{\rightarrow}

\usepackage{rotating}
\usepackage{etoolbox} 
\usepackage{courier}
\usepackage{empheq}
\newsavebox\CBox
\def\textBF#1{\sbox\CBox{#1}\resizebox{\wd\CBox}{\ht\CBox}{\textcolor{black}{\textbf{#1}}}}

\newcommand{\mcom}[1]{\textbf{\texttt{#1}}}
\newcommand{\mtext}[1]{\textbf{\texttt{#1}}}
\newcommand{\mout}[1]{\texttt{#1}}

\patchcmd{\subsubsection}{\itshape}{\bfseries}{}{}

\newcommand{\Block}[1] {\small \textcolor{mygreen}{ \fontfamily{phv}\selectfont{{In[#1]}}}}
\newcommand{\Blocko}[1] {\small \textcolor{mygreen}{ \fontfamily{phv}\selectfont{{Out[#1]}}}}
\definecolor{mygreen}{rgb}{0.204, 0.408, 0.408}

\begin{document}

\title{Solving the One-Dimensional Time-Independent Schr\"odinger Equation with High Accuracy: The \mtext{LagrangeMesh} Mathematica$^\circledR$ Package}

\author{J.C. del Valle}
	\email{juan.delvalle@ug.edu.pl}

\affiliation{Institute of Mathematics, Faculty of Mathematics, Physics, and Informatics, University of Gda\'nsk, 80-308 Gda\'nsk, Poland}

\begin{abstract}
In order to  find the spectrum associated with the one-dimensional Schr\"odinger equation, we discuss the Lagrange Mesh Method (LMM)  and its numerical implementation. After presenting a general overview of the theory behind the LMM, we introduce the \mcom{LagrangeMesh} package: the numerical implementation of the LMM in  Mathematica$^{\circledR}$. Using few lines of code, the package enables a quick home-computer and highly accurate computation of the spectrum  and provides a practical tool to study  a large class of systems in quantum mechanics. The main properties of the package are (i) the input is the potential function and the interval on which it is defined; and (ii) the accuracy in calculations and final results is controllable by the user.   Due to its high accuracy and simple usage,  the package may be used as a research and educational tool.
As  illustration, a highly accurate  spectrum of some relevant quantum systems  is obtained by employing the commands that the package offers.
\end{abstract}

\maketitle
	\newpage

\section{Introduction}
Solutions  to  the time-independent Schr\"odinger equation \cite{Schrodinger1926} are crucial for our current understanding of quantum mechanics. 
However, only a handful of systems described by this equation admit exact solutions. Since the early days of quantum mechanics, the lack of exact solutions  has stimulated the development of  methods to find them in approximate form.  As a result,    nowadays we have a wide variety of  accurate numerical approaches to solve the Schr\"odinger equation.

In this paper, we describe and implement computationally one of such methods outstanding for its simplicity: the Lagrange Mesh\cite{Baye2015}. In general terms, the Lagrange Mesh Method (LMM) is an approximate variational approach in which a linear combination of normalizable \textit{Lagrange functions} approximates the exact eigenfunctions.  These  functions are related to a set of mesh points and  the Gauss quadrature associated with it. Thus, it belongs to the family of \textit{pseudospectral} methods. The input needed by the method is minimal: the potential and the domain in which is defined. The outcome is the lowest eigenvalues and eigenfunctions in approximate numerical and analytical \textcolor{black}{forms}\footnote{ \textcolor{black}{Analytical form is only available for  eigenfunctions.}}.  

The main object of the present work is the one-dimensional time independent Schr\"odinger equation,
\begin{equation}
	-\frac{\hbar^2}{2m}\pa_x^2\psi(x)\ +\ V(x)\psi(x)\ = \ E\,\psi(x)\ ,\qquad \pa_x\equiv\frac{d}{dx}\ ,
	\label{Schrodinger}
\end{equation}
defined on some interval $(a,b)$. We assume that the  potential $V(x)$ is \textcolor{black}{a smooth} function,  not necessarily real,  and that any eigenfunction vanishes at the endpoints.    For equation (\ref{Schrodinger}), we describe and  implement the LMM computationally  for all possible  intervals of the form $(a,b)$: finite, semi-infinite, and infinite. Under certain conditions  described in this paper, our implementation of the  LMM leads to a highly accurate spectrum obtained in short CPU times.
Furthermore,  since the potential $V(x)$ is not assumed to be real,  resonance states can be studied using the complex scaling  \textit{technique}, see Ref. \cite{Simon2}. In addition,  $\mathcal{PT}$-symmetric potentials can be studied as well.
At this point, it is worth mentioning that the one-dimensional Schr\"odinger equation (\ref{Schrodinger}) also may used to study  $d$-dimensional radial potentials, usually defined in $(0,b)$.   In such cases, an \textit{effective} potential must be considered as $V(x)$. Thus, the present implementation can  tackle such multidimensional cases.

There are systems for which a highly accurate  solution of (\ref{Schrodinger}) is  needed in a natural way.  For example, the calculation of critical  parameters \cite{Simon} of a  confining potential\footnote{The potential is assumed to be defined on an (semi)infinite domain, and with the property $\lim\limits_{|x|\rar\infty}V(x;\la)\rar0$.} $V(x;\la)$ that depends on some parameter $\la$. Assuming that  $\la$ controls  either  the depth or the width of the potential, critical phenomena may occur, see \textcolor{black}{Ref. \cite{Simon}}. If so,  for given bound  eigenstate  $\ket{n}$, there exists a certain value of $\la$, namely the critical  parameter $\la_c(n)$, at which the  energy $E_n$ is \textit{absorbed} into the continuous spectrum and $\ket{n}$ is no longer bound. 	Typically, eigenfunctions are usually extremely flat and extended in the domain if $\la$ is close to $\la_c$. These features make the calculation of critical parameters a task for which  traditional methods\footnote{See Ref. \cite{Yukawa} and references therein for a concrete example based on the Yukawa potential.} frequently fail due to their limited accuracy (e.g., those based on finite differences, Runge-Kutta, perturbation theory, etc.). Furthermore, while other methods fail, \textcolor{black}{pseudospectral} methods have shown to be  adequate for studying critical behavior, see Ref. \cite{Yukawa}. 

Another example comes from  the field of  non-perturbative  effects in quantum mechanics. For example,  when $V(x)$ is  a multi-well potential with degenerate minima those effects are reflected on the spectrum. 
The prime example  is the polynomial quartic double-well\cite{Zinn}:  $V(x)=\frac{1}{2}x^2(1-gx)^2$.  At $g=0$, $V(x)$ has only one minimum located at $x=0$ and describes a harmonic oscillator. In particular, its ground state is characterized by energy $E_0=-1/2$. If we consider $g\neq0$, the ground state energy splits into two: $E^{+}$ and $E^{-}$. They are separated by an exponentially small energy gap when $g$ is small. It reflects the non-perturbative effects on the eigenvalues. The separation can be calculated in the form of a trans-series by means of instanton contributions from a semi-classical calculation based on the path integral formalism \cite{Zinn}. Nevertheless, from a practical point of view, the numerical calculation of such small gaps can be done straightforwardly by the LMM thanks to its high accuracy. On the other hand, recent advances on the field  \cite{Dunne}  have  shown a sharp contrast between the double-well and the multi-well potentials while developing a semi-classical consideration based on instantons. In particular, for the triple-well polynomial potential $V(x)=\frac{1}{2}x^2(1-gx^2)^2$, the low-lying energy levels for states localized in the outer wells are exponentially split, while the energy	levels for states localized in the inner well are not split at all. Those non-perturbative small effects on eigenvalues  for an arbitrary multi-well potential can be captured by the LMM. For these kind\textcolor{black}{s} of computations,  using arbitrary-precision is mandatory.

The aim of the present paper is two-fold.  From one side, we give a concrete discussion of the LMM, which can serve as a pedagogical  introduction to the method itself. From the other, we introduce the \mtext{LagrangeMesh} package: the numerical implementation of the LMM  written in Mathematica$^\circledR$ 13. Once installed, it will provide the user with five commands to realize the LMM on equation (\ref{Schrodinger}).   
We chose this widely used programming language due to its clearness and compactness while coding, as well as for its forward compatibility with newer versions. Furthermore, we exploit the arbitrary precision arithmetic that Mathematica$^\circledR$ provides by simply specifying  the option \mtext{WorkingPrecision}. In fact, all commands supplied by the package \mtext{LagrangeMesh} are equipped with this option to control the loss of accuracy during calculations.
It allows the user to reach and overcome benchmarks for eigenvalues and eigenfunctions  in a few lines of code using a standard nowadays computer. Therefore, the package is  user-friendly.  Compared with  previous implementations\footnote{Most of them written in FORTRAN  working with double  and, sometimes, quadruple precision. See \textcolor{black}{Ref. \cite{Baye2015}} and references therein. }, these characteristics make the \mtext{LagrangeMesh} package superior in terms of  accuracy control. Furthermore,  Mathematica$^\circledR$ already counts with alert messages estimating errors, and  the loss of accuracy during calculations. This feature is fundamental while obtaining \textcolor{black}{highly accurate} results (e.g. critical parameters). All those built-in error estimates  are inherited directly to the package.

In the last couple of years,  efficient numerical  solvers of (\ref{Schrodinger}) have been developed in different programming languages: C++ and Python \cite{BAEYENS}, Fortran \cite{Salvat},  Mathematica \cite{Atkin},  C and Matlab \cite{Barrio}. They work with double and quadrupole precision. It means, for example, that non-perturbative effect in the weak coupling regime, and the estimation of accurate critical parameters is out of their scope. As above-mentioned,   \mtext{LagrangeMesh} works with arbitrary precision and does not have these drawbacks and, as a result, it  complements such solvers. 

The present work is organized as follows: in  Sections  \ref{1DLMM},  \ref{Formulas}, and \ref{Transformations}, we present a description of the LMM. The discussion is aimed to give the user the fundamental concepts behind the package. Undoubtedly, knowing the niceties and limitations of the method may lead to  better performance during calculations. 

 In  Section \ref{TheLMMPackage},  we introduce the \mtext{LagrangeMesh} package with a detailed description of the usage and limitations. 
Actually, this Section can be regarded as a user guide based on worked examples.  Specifically, we complement the discussion with concrete applications. First, we show how commands are used for obtaining the spectrum of exactly solvable potentials. These examples set the ground for the study of relevant  quantum-mechanical systems/potentials: quartic anharmonic oscillator, quartic double-well with degenerate minima,  shell-confined hydrogen atom, $\mathcal{P}\mathcal{T}$-symmetric cubic oscillator, quasi-exactly solvable sextic anharmonic potential, and  Rydberg atoms.
Those worked examples allow us to study different phenomena that appear in the quantum spectrum and glimpse the scope of the package. For instance, for the  quartic double well, we can calculate the exponentially small gap between the two lowest states and study the  partial sums of the semi-classical resurgent expansion \cite{Zinn}. In general, the results obtained by the package  are compared with those found in the literature.
To simplify Section \ref{TheLMMPackage}, we present \textit{blocks of codes} that show concrete numerical implementations. They display the input and output in the same way \textcolor{black}{that} Mathematica$^\circledR$ does. In addition, we show some plots generated  in Mathematica$^\circledR$ and their corresponding codes from which they were generated. We have avoided labeling the axis of  plots to keep blocks as simple and compact as possible.

\section{The One-Dimensional  Lagrange Mesh Method }
\label{1DLMM}
Three basic ingredients are the building blocks of the LMM:  the Gauss quadrature approximation, Lagrange functions, and the secular equations. In the following Sections, we give an overview of each one. Full details  can be found in \cite{Baye2015} and references therein. 

\subsection{Gauss Quadrature}
Consider the one-dimensional integral over a smooth function $f(x)$ on the interval\footnote{The interval of integration can be finite $[a,b]$, semi-infinite $[a,\infty)$ or $(-\infty,b]$,  or infinite $(-\infty,\infty)$.}  $[a,b]$, namely,
\begin{equation}
	\int^{b}_a f(x)\,w(x)\,dx\ ,
	\label{basicI}
\end{equation}
where $w(x) \geq 0$ is a given weight function. Frequently  this integral can not be computed  analytically. Nevertheless, numerical integration methods can circumvent this inconvenience. Among those methods, the Gauss quadrature approximation stands alone due to its simplicity, efficiency, and high accuracy. The basic idea of this approach is to approximate the integral (\ref{basicI}) by a particular finite sum:
\begin{equation}
	\int^{b}_a f(x)\,w(x)\,dx\ \approx\ \sum_{k=1}^{N}w_kf(x_k)\ .
	\label{quadrature}
\end{equation}
The set $\{x_k\}_{k=1}^{N}$ contains the \textit{mesh points}, while  $\{w_k\}_{k=1}^{N}$ \textcolor{black}{are}  the (positive) \textit{weights}. The expression shown on the right-hand side of (\ref{quadrature}) is the so-called  \textit{ quadrature approximation}.

To determine $\{x_k\}_{k=1}^{N}$  and $\{w_k\}_{k=1}^{N}$,  we demand the   quadrature approximation
to be exact when the function $f(x)$ is a polynomial of the degree $(2N-1)$. As a result, it can be shown that   $\{x_k\}_{k=1}^{N}$  and $\{w_k\}_{k=1}^{N}$ are real and they may be obtained by  solving  algebraic equations. However, there is a more efficient way to find mesh points and weights in practice. It can be demonstrated \cite{Atkinson} that the $N$ mesh points $\{x_k\}_{k=1}^{N}$ are nothing but the $N$ zeroes of the $N$-degree orthogonal polynomial $\mathcal{P}_N(x)$ associated with the weight function\footnote{For a given weight function $w(x)$, one can always construct a set of orthogonal polynomials $\{\mathcal{P}_k(x)\}_{k=1}^N$using the \text{Gram-Schmidt}  process.}. In addition, any $w_k$ is basically given in terms of $x_k $, $\mathcal{P}_{N-1}(x_k)$, and $\mathcal{P}_N'(x_k)$, see \cite{Baye2015}. However, the well-known explicit form of the weights is irrelevant for the LMM, so we omit to present it. 

Once we choose  $[a,b]$ and  $w(x)$, $\{x_k\}_{k=1}^{N}$  and $\{w_k\}_{k=1}^{N}$  are completely defined.
For example, taking $w(x)=1$ and the interval $[-1,1]$, the mesh points are the $N$ zeroes of the Legendre polynomial  of \textcolor{black}{degree} $N$. The name of the mesh is usually given according to the name of the polynomials involved. Therefore, in the  above example, $\{x_k\}_{k=1}^{N}$ constitutes a {\it Legendre} mesh of $N$-points.

It is worth noting that, by construction, the accuracy of the Gauss quadrature approximation (\ref{quadrature}) depends on how well $f(x)$ is approximated by a polynomial of  degree $(2N-1)$ in $[a,b]$.  If $f(x)$ is $2N$-times differentiable, the error estimate is given by
\begin{equation}
	\left|\int_{a}^bf(x)w(x)\,dx\ -\ \sum_{k=1}^Nw_kf(x_k)\right|\ \leq\ \frac{\int_a^b\mathcal{P}_N(x)^2\,dx}{(2N)!}\,\underset{\xi\in[a,b]}{\sup}\{f^{(2N)}(\xi)\}\ ,
\end{equation} 
see Ref. \cite{Atkinson} for details and derivation. Consequently, if the function $f(x)$ is sufficiently smooth, the accuracy of  the Gauss quadrature approximation usually increases as $N$ does. However, it may fail if  $f(x)$ has singularities, discontinuities, or it is not differentiable at some points inside $[a,b]$. 

In practice, the appearance of the weight function $w(x)$ in the integrand of (\ref{basicI}) can be omitted by defining an auxiliary function
\begin{equation}
	g(x)\ =\ f(x)\,w(x)\ .
	\label{hide}
\end{equation}
In this way, the Gauss quadrature approximation (\ref{basicI}) now takes the form
\begin{equation}
	\int^{b}_a g(x)\,dx\ \approx\ \sum_{k=1}^{N}\la_kg(x_k)\ , \qquad 	\la_k\ =\ \frac{w_k}{w(x_k)}\ ,\qquad k=1,2,...,N\ .
	\label{quadrature2}
\end{equation}
Now $\{\lambda_k\}_{k=1}^{N}$ play the role of weights. Throughout this paper, we will work with the Gauss quadrature approximation  defined according to (\ref{hide}) and (\ref{quadrature2}). In addition, for the purposes of the present work, only meshes and weights associated with  Legendre, Laguerre, and Hermite classical orthogonal polynomials are relevant. 

\subsection{Lagrange Functions}
The second ingredient of the LMM is a special set of $N$  smooth (infinitely differentiable) real functions $\{f_i(x)\}_{i=1}^N$, which are called Lagrange functions. To define them, first we need to choose a particular mesh $\{x_k\}_{k=1}^{N}$ and know the corresponding weights $\{\la_k\}_{k=1}^N$.   By definition, a Lagrange function should satisfy two requirements:
\begin{enumerate}
	\item The Lagrange condition, which is
	\begin{equation}
		f_{i}(x_j)\ =\ \la_i^{-1/2}\delta_{ij} \ ,\qquad i,j\ =\ 1,2,...,N\ .
		\label{p1}
	\end{equation}	
	\item  Lagrange functions  are orthonormal in the Gauss quadrature approximation (GQA), namely
	\begin{equation}
		\int_{a}^{b}f_{i}(x)f_j(x)\,dx\ \overset{\text{GQA}}{=}\ \delta_{ij} \ .
		\label{p2}
	\end{equation}
\end{enumerate}

There are well-known prescriptions to construct Lagrange functions based on orthogonal polynomials. \textcolor{black}{They can be found explicitly in Refs. \cite{Baye2015} and  \cite{Schwartz}, so we skip  details.}  For the cases of our interest (Legendre, Laguerre, Hermite), explicit formulas of Lagrange functions are  presented further down. Since the explicit form of the weights \textcolor{black}{$\{\la_k\}_{k=1}^{N}$} can be extracted from the Lagrange condition (\ref{p1}), we avoid presenting such expressions.

\subsection{The Secular Equations and Lagrange Mesh Equations}

The third ingredient of the LMM is the set of the so-called secular equations.  These equations are useful for finding in approximate form the spectrum of a given  time-independent Schr\"odinger equation,
\begin{equation}
	\hat{H}\psi\ =\ E\,\psi\ ,
	\label{SEq}
\end{equation}
where $\psi$ is the wavefunction (eigenfunction), $E$ the energy (eigenvalue), and  $\hat{H}$ the Hamiltonian.\footnote{The terms eigenvalue and energy will be used indistinctly, a similar situation with eigenfunction and wavefunction.} In particular, we are interested in the  one-dimensional Hamiltonian of the form
\begin{equation}
	\hat{H}\ =\ -\frac{\hbar^2}{2m}\pa_x^2\ +\ V(x)\ ,
	\label{SH}
\end{equation}
where $m$ is the mass of the particle, $V(x)$ a  confining potential\footnote{\textcolor{black}{A potential for which the Schr\"odinger operator may have a discrete spectrum bounded from below whose corresponding wavefunctions are normalizable in the sense of (\ref{norm}).}}, and $\hbar$ the reduced Planck constant. We assume that $\hat{H}$ is defined in $(a,b)$, and  the boundary conditions imposed on eigenfunctions are
\begin{equation}
	\psi(a)\ =\ \psi(b)\ =\ 0 \ .
	\label{Boundary}
\end{equation}
Under the above-mentioned conditions, the spectrum of the Hamiltonian (\ref{SH}) is real and discrete for the so-called \textit{bound states}. These states are characterized by a square-integrable eigenfunction, such that
\begin{equation}
	\int_a^b|\psi(x)|^2\,dx\ =\ 1\ .
	\label{norm}
\end{equation}

As we previously mentioned, frequently the spectral problem defined by (\ref{SEq}) cannot be   solved in exact form. To find an approximate solution, let us suppose  that any wavefunction can be approximated by means of the linear combination\cite{Schwartz}
\begin{equation}
	\psi(x)\ \approx\ \sum_{k=1}^{N}c_k\,\phi_k(x)\ .
	\label{approxWF}
\end{equation}
Here \{$\phi_k\}_{k=1}^{N}$ is a set of $N$ orthonormal functions\footnote{They are assumed to be real for simplicity.} which are chosen in such a way that boundary conditions (\ref{Boundary}) are fulfilled.  In turn, coefficients $c_k$ are unknown   for the moment\textcolor{black}{, but they satisfy
\begin{equation}
	\sum_{k=1}|c_k|^2\ =\ 1\ 
\end{equation}
in order to fulfill (\ref{norm})}. Using the variational principle, one can easily show that  $c_k$ are determined by the equations
\begin{equation}
	\sum_{j=1}^{N}\left(\frac{\hbar^2}{2m}T_{ij}\ +\ V_{ij} \right)c_j\ =\ E\,c_i\ ,\qquad i=1,2,...N\ ,
	\label{secular}
\end{equation}
where
\begin{equation}
	T_{ij}\ =\ -\int_{a}^{b} \phi_i(x)\,\pa_x^2\phi_j(x) \,dx\ ,\qquad V_{ij}\ =\ \int_{a}^{b} \phi_i(x)V(x)\phi_j(x) \,dx\ .
	\label{elements}
\end{equation}
Equations (\ref{secular}) are the so-called \textit{secular equations}, see Ref. \cite{Landau} for details.  They establish that, in order to find $c_k$, we have to diagonalize the matrix representation of the Hamiltonian constructed with the functions \{$\phi_k\}_{k=1}^{N}$. As a result, eigenvalues of this matrix correspond to approximations of the  exact energies; meanwhile, eigenvectors contain coefficients $c_k$ that ultimately \textcolor{black}{determine the approximate  wavefunctions.} 

Note that  $T_{ij}$ and $V_{ij}$ in (\ref{elements}) play the usual role of kinetic and potential matrix elements, respectively.
In the LMM, those matrix elements are calculated by taking $\phi_k(x)$ as Lagrange functions $f_k(x)$  and using the Gauss quadrature to compute the integrals involved  in matrix elements. This approach connects the three ingredients discussed above. Under these considerations, the approximate secular equations  (\ref{secular}) now read
\begin{equation}
	\sum_{j=1}^{N}\left(\frac{\hbar^2}{2m}T_{ij}^{(G)}\ +\ V_{ij}^{(G)} \right)c_j\ =\ E\,c_i\ ,\qquad i=1,2,...N\ ,
	\label{lagrange_eqs}
\end{equation}
where the super-index $(G)$ indicates that we are using the Gauss quadrature  when integrating (\ref{elements}). In this sense, due to the Gauss quadrature approximation, the LMM is an approximate variational method. Equations presented in  (\ref{lagrange_eqs}) are called \textit{Lagrange equations}. As a consequence of using Lagrange functions and the Gauss quadrature approximation, the potential matrix elements are diagonal,
\begin{equation}
	V^{(G)}_{ij}\ =\ V(x_i)\,\delta_{ij}\ .
	\label{VG}
\end{equation}
Hence, its computation is straightforward: the potential is evaluated at the mesh points. This property makes the LMM a versatile approach to study, in principle,  any confining potential. The only requirement is  that (\ref{VG}) approximates with high accuracy the exact potential elements shown in (\ref{elements}). On the other hand, the kinetic matrix elements $T_{ij}^{(G)}$ acquire simple and compact expressions written in terms of $\{x_i,x_j\}$  and $N$. Explicit formulas for relevant cases (Legendre, Laguerre, and Hermite) are presented below.

To conclude this Section, let us indicate an important remark concerning the calculation of the  expectation value of a given  scalar operator $\hat{O}=O(x)$. In the context of the LMM, the expectation value of $\hat{O}$,
\begin{equation}
	\braket{\hat{O}}\ =\ \int_{a}^b\psi^*(x)O(x)\psi(x)\,dx\ ,
\end{equation}	
is calculated taking into account the Gauss quadrature approximation when integrating. Since $\psi(x)$ is a linear combination of Lagrange functions, it is easy to see that
\begin{equation}
	\braket{\hat{O}}_{(G)}\ =\ \sum_{k=1}^{N}|c_k|^2O(x_k)\ ,
	\label{expectation}
\end{equation}	
where the sub-index $(G)$ indicates the usage of the Gauss quadrature. Equation (\ref{expectation}) is a direct consequence of the properties  (\ref{p1}) and (\ref{p2}). In summary, the calculation of expectation values of scalar operators is reduced to calculate the sum (\ref{expectation}).

\section{Formulas for Lagrange Functions and Kinetic Elements}
\label{Formulas}
Now we present  three  Lagrange functions defined for finite, semi-infinite, and infinite intervals. They all are based on the following classical orthogonal polynomials: Legendre, Laguerre, and Hermite, respectively. For completeness, explicit formulas of the  kinetic matrix elements  in the Gauss quadrature  approximation $T_{ij}^{(G)}$ are  shown. As we will see, they are written in compact expressions.  Details concerning their derivation can be found in \textcolor{black}{Ref. \cite{Baye2015} }and references therein.

\subsection{Finite Domain}
For the finite interval $[-1,1]$, it is natural to use a  Legendre mesh. The corresponding Lagrange functions read
\begin{equation}
	f_i(x)\ =\ \frac{\textcolor{black}{(-1)^{n+i}} (x+1) (1-x)}{\sqrt{2(x_i+1) (1-x_i)}}\,\frac{P_N(x)}{x-x_i} \ ,\qquad i\ =\ 1,2,...,N\ .
	\label{Legendre}
\end{equation}
Here $P_N(x)$ denotes the $N$-th Legendre polynomial.
By construction, any Lagrange function $f_i(x)$ satisfies   $f_i(-1)=f_i(1)=0$.  Thus, boundary condition (\ref{Boundary}) is fulfilled for $a=-1$ and $b=1$.  The kinetic  matrix elements $T_{ij}^{(G)}$ are given by   
\begin{equation}
	T_{i\neq j}^{(G)}\ =\ \textcolor{black}{\frac{(-1)^{i+j+1} ( 2x_i x_j-2)}{(x_i-x_j)^2 \sqrt{(1-x_i^2)(1-x_j^2)}}}\ ,\qquad
	T_{ii}^{(G)}\ =\ \frac{N(N+1)  (1-x_i^2)+4}{3  (1-x_i^2)^2} \ .
	\label{LegendreKinetic}
\end{equation}
Functions shown in (\ref{Legendre}) are building blocks to consider arbitrary but finite, intervals
of the form $[a,b]$. This can be achieved via a monotonic mapping $t:[-1,1]\rar[a,b]$, see below for discussion. 

\subsection{Semi-Infinite Domain}
Now we consider  $[0,\infty)$.  In this case, the natural mesh is a Laguerre one. The corresponding Lagrange functions are
\begin{equation}
	f_i(x)\ =\ (-1)^{N+i}\frac{x}{\sqrt{x_i}}\,\frac{  L_N(x)}{x-x_i}\,e^{-x/2}\ ,\qquad i\ =\ 
	1,2,...N\ .	\label{Laguerre}
\end{equation}
Here $L_N(x)$ denotes the $N$-th Laguerre polynomial. 
From (\ref{Laguerre}), it is clear that any Lagrange function $f_i(x)$ satisfies $f_i(0)=f_i(\infty)=0$. Therefore, boundary condition (\ref{Boundary}) is fulfilled for $a=0$ and $b=\infty$.  Explicit formulas for kinetic  matrix elements $T_{ij}^{(G)}$ are 
\begin{equation}
	T_{i\neq j}^{(G)}\ =\ 	\frac{(-1)^{i-j} (x_i+x_j)}{\sqrt{x_i \,x_j} (x_i-x_j)^2}\ ,\qquad T_{ii}^{(G)}\ =\ 	\frac{(4 N+2) x_i-x_i^2+4}{12\,x_i^2}\ .
\end{equation}
Note that with a global translation of mesh points and Lagrange functions (\ref{Laguerre}), we can tackle domains of the form $[a,\infty)$. 
Another important remark concerning functions (\ref{Laguerre}) must be emphasized. For now, consider the reduced three-dimensional radial Schr\"odinger Hamiltonian
\begin{equation}
	\hat{H}\ =\ -\frac{\hbar^2}{2m}\pa_r^2\ +\ \frac{\hbar^2\ell(\ell+1)}{2mr^2}+\ V(r)\ ,\qquad\ 0\ \leq r\ <\infty\ ,\qquad \ell\ =\ 0,1, ...
	\label{rSEq}
\end{equation}
where $r$ denotes the radial variable in the standard spherical coordinate system, and $\ell$ the angular momentum quantum number. If the radial potential $V(r)$ is confining, the boundary conditions
\begin{equation}
	\psi(0)=\ 0\ ,\qquad \psi(\infty)\ =\ 0\ 
\end{equation} 
\textcolor{black}{together with (\ref{norm})} are usually imposed when solving the reduced radial Schr\"odinger equation $\hat{H}\psi(r)=E\psi(r)$. Hence, Lagrange functions (\ref{Laguerre}) are adequate to approximate  three-dimensional radial eigenfunctions in the form of (\ref{approxWF}).  A similar situation occurs with  Lagrange functions based on Legendre polynomials (\ref{Legendre}): they are useful to study radial potentials when the particle is confined to a spherical core-shell or shell-type cavity. Explicit examples will be presented further in the text. Using similar arguments, (\ref{Laguerre}) are suitable to tackle any $d$-dimensional reduced radial Schr\"odinger equation.  \textcolor{black} {However, due to boundary conditions, the case $d=2$ must be excluded for $S$-states.}

\subsection{Infinite Domain}
For the one-dimensional case in $(-\infty,\infty)$,  it is natural to consider the Hermite mesh. The corresponding Lagrange functions,  read
\begin{equation}
	f_i(x)\ =\ \frac{(-1)^{N-i}}{(2\,h_N)^{1/2}}\,\frac{H_{N}(x)}{x-x_i}\,e^{-x^2/2}\ ,\qquad h_N\ =\ 2^NN!\sqrt{\pi}\ ,\qquad i=1,2,...,N\ ,
	\label{hermite}
\end{equation}
where $H_N(x)$ denotes the $N$-th Hermite polynomial.
It is clear that any $f_i(x)$ satisfies $f_i(-\infty)=f_i(\infty)=0$, since it decays exponentially at large $|x|$.  Therefore, boundary condition (\ref{Boundary}) is fulfilled for $a=-\infty$ and $b=\infty$. Kinetic matrix elements $T_{ij}^{(G)}$ are 
\begin{equation}
	T_{i\neq j}^{(G)}\ =\ (-1)^{i-j}\,\frac{2}{(x_i-x_j)^2}\ ,\qquad T_{ii}^{(G)}\ =\ \frac{1}{3}(2N+1-x_i^2)\ .
\end{equation}
\section{Discretization,  Scaling, and Mapping}
\label{Transformations}
Before introducing the \mtext{LagrangeMesh} package, there are some aspects of the method that have to be discussed. In particular, they play an essential role when performing  a highly accurate calculation of the spectrum.  
\subsection{Discretization of the Eigenfunctions}
In the framework of the LMM, the approximate wavefunction has the representation
\begin{equation}
	\psi(x)\ =\ \sum_{k=1}^{N}c_k\,f_k(x)\ , \qquad a \leq x \leq b\ ,
	\label{Continuos}
\end{equation}
where $f_k(x)$ is the $k$-th Lagrange function. 
Note that if $N$ is large, the number of arithmetic operations involved when  evaluating  (\ref{Continuos}) numerically at some arbitrary point $x$ can be huge. Therefore, the accumulation of error can play an important role  and it might lead to wrong results for $\psi(x)$.  There is a way to avoid such situation, but we have to abandon the idea of constructing a continuous wavefunction like  (\ref{Continuos}). Through the Lagrange condition (\ref{p1}), a $N$-point discrete version of $\psi(x)$ defined at the non-uniform mesh points $\{x_k\}_{k=1}^N$ can be obtained by noting that
\begin{equation}
	\psi(x_k)\ =\ \frac{c_k}{\lambda_k^{1/2}}\ ,\qquad k=1,2,...,N\ .
	\label{discrete}
\end{equation}
This discrete representation usually leads to better approximations of $\psi(x_k)$ than evaluating directly via (\ref{Continuos}).

\subsection{Scaling} 
Higher accuracy in results predicted by the  LMM can be obtained if the  mesh points lie inside the region where the wavefunction is not \textit{too small}. A global scaling of the mesh points in the form
\begin{equation}
	x_k \rar  h\,x_k\ ,\qquad h>0\ ,\qquad  k=1,2,...,N \ ,	
	\label{scaling}
\end{equation}
may move them to such region. We have denoted  the global scaling parameter by $h$. If it is chosen appropriately, it can increase accuracy and improve the performance of the LMM  reducing CPU times.
In addition, Lagrange functions should be modified accordingly, 
\begin{equation}
	f_i(x)\rar \frac{1}{h^{1/2}}f_i(x/h)\ .  
\end{equation}
The factor $h^{1/2}$ is introduced to fulfill the Lagrange and \textcolor{black}{ orthonormality conditions: (\ref{p1}) and (\ref{p2}), respectively}. 
The usage of $h$ not  only modifies the mesh points and the Lagrange functions,  it also affects the Lagrange equations.  They now read 
\begin{equation}
	\sum_{j=1}^{N}\left(\frac{\hbar^2}{2mh^2}T_{ij}^{(G)}\ +\ V(hx_i)\delta_{ij} \right)c_j\ =\ E\,c_i\ ,\qquad i=1,2,...,N\ .
	\label{LMM_eq_h}
\end{equation}
Sometimes  $h$ can  be used as a variational parameter, especially if $N$ is large. However, the Gauss quadrature approximation used in (\ref{LMM_eq_h}) may lead to a fake minimum in energy with respect to $h$.

\subsection{ Mapping}
The above-mentioned global scaling transformation of a mesh is one of the simplest mappings that can be implemented to move the mesh points. Let us focus on the Legendre mesh and the corresponding Lagrange functions. This mesh is originally defined in $[-1,1]$ by construction, but it can be easily mapped into $[a,b]$ using the following linear function \cite{Castano},
\begin{equation}
	t(x)\ =\ \frac{b-a}{2}x \ +\ \frac{a+b}{2}\ ,\qquad t(-1)\ =\ a\ ,\qquad t(1)\ =\ b\ .
\end{equation}
This simple function allows us to scale, translate, reflect, and map the mesh points. In this way, we can   realize the LMM in an arbitrary finite domain apart from $[-1,1]$. Lagrange functions are modified as well according to
\begin{equation}
	f_i(x)\rar \sqrt{\frac{2}{|b-a|}}\,f_i(t^{-1}(x))\ .  
\end{equation}
Similar considerations can be made for the Laguerre mesh. Instead of providing further details, which can be found in \textcolor{black}{Ref. \cite{Baye2015}}, we point out that all necessary transformations are implemented in the \mtext{LagrangeMesh} package internally and automatically. In this way, the user can study spectral problems defined in any one-dimensional interval.

\section{The \mtext{LagrangeMesh} Mathematica$^\circledR$ package}
\label{TheLMMPackage}
The \mtext{LagrangeMesh} package\footnote{Designed in Mathematica$^\circledR$ 13, and tested in versions  12 and 13.} implements the LMM for solving the one-dimensional Schr\"odinger equation in different domains for an arbitrary potential.

While presenting the main characteristics of the package, we focus on working out  particular examples to show the usage. As we mentioned previously, we use  blocks of codes that display the input, output, and syntax of each command.  \textcolor{black}{All calculations were obtained in a Mac mini (late 2014), 2.6 GHz and 8GB. }
All worked examples can be found in the supplemental file called \mtext{WorkedExamples.nb}. The  reader without previous experience with the LMM is encouraged to refer to it. Throughout the rest of this text, the value of the reduced Planck constant $\hbar$ is  set to the unity $(\hbar=1)$. In fact, the package is written with this normalization for $\hbar$.

\subsection{Installation}
The  file called \mtext{LagrangeMesh.wl} contains the package. Once downloaded, there are two ways to use it.

\begin{enumerate}
	\item Full installation of the package. This can be achieved via the Install option located at the menu of Mathematica$^\circledR$: File$\rar$Install. The procedure is straightforward, so we omit details that can be found in \textcolor{black}{Ref.} \cite{InstallMathematica}. Once the installation was successful, the package is loaded evaluating \mtext{Needs["\textcolor{gray}{LagrangeMesh`}"]} in the notebook we want to use. Each time the kernel is restarted, we need to load the package as indicated. 
	
	\item Loading the package temporarily. This is a simple alternative that  requires no installation. This option only works  if the package and the notebook (in which we will perform calculations) are contained in the same directory. Loading the package is achieved by evaluating \mtext{<<\textcolor{gray}{"path/LagrangeMesh.wl}"} in the notebook file we want to work with. The explicit form of \mtext{\textcolor{gray}{path}} is found evaluating the command \mtext{NotebookDirectory[]}.   Each time the kernel is restarted, we need to load the package as indicated. 
\end{enumerate}	

\subsection{Commands}
Once installed or loaded, the \mtext{LagrangeMesh} package will provide the user five new commands. Two related to the construction of meshes and weights: \mcom{BuildMesh} and  \mcom{AvailableMeshQ}; and three additional for realizing the LMM: 
\mcom{LagMeshEigenvalues},  \mcom{LagMeshEigenfunctions}, and  \mcom{LagMeshEigensystem}.
These commands are designed with the standard Mathematica$^\circledR$ style (coloring, syntax, etc.). In addition, each command counts with several alert messages that can guide the user to correct  syntax errors. Below, we describe the general use and syntax of each command. 

\subsubsection{\mcom{BuildMesh}}
\begin{figure*}[h]
	\centering
	\includegraphics[scale=0.9]{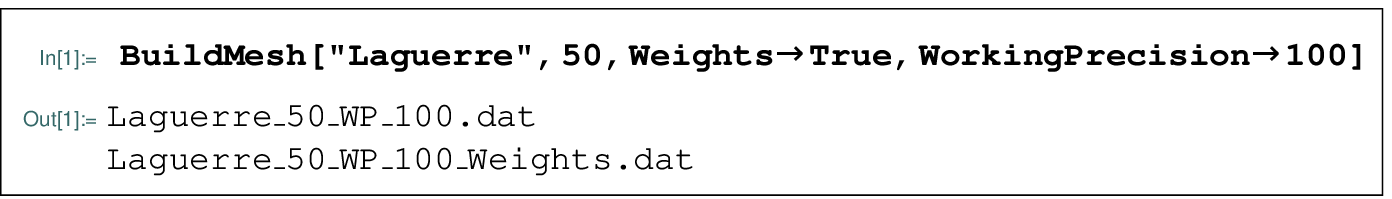}
	\label{B1}
\end{figure*}

This command constructs  mesh points (zeroes) and weights for a given type of classical orthogonal polynomial. The syntax of this command is the following:
\begin{equation*}
	\small
	\mcom{BuildMesh[Type,Dimension,Options]}
\end{equation*}
\begin{itemize}[label=$\bullet$,leftmargin=*]
	\item	\mcom{Type} specifies the classical polynomial, it can only take the value of the following strings \textcolor{gray}{\mtext{"Legendre"}},  \textcolor{gray}{\mtext{"Laguerre"}}, or   \textcolor{gray}{\mtext{"Hermite"}}. 
	\item \mcom{Dimension} defines the number of mesh points. Therefore, its value must be a positive  integer number. 
	
	\item There are two \mtext{Options} that can be specified for  \mcom{BuildMesh}. They allow the user to compute  the Gaussian weights and to control the accuracy of calculations. Each option is described in Table \ref{Options:BuildMesh}
	
\begin{table}[ht]
\caption{Options for the command \mcom{BuildMesh}.}
 \resizebox{\textwidth}{!}{	{\begin{tabular}{@{}clc@{}} \toprule
		\mcom{Option}& Description & Default Value  \\ \colrule
			\mcom{Weights} & If the value is set to \mtext{True}, the  weights associated & \mcom{False}  \\
			&to the mesh will be calculated.  \\ \\
			\mcom{WorkingPrecision} &It specifies how many digits of precision should be   & \mcom{MachinePrecision} \\ 
			&maintained in internal computations.\\ 
			
			\botrule
		\end{tabular} \label{ta1}}}
			\label{Options:BuildMesh}
\end{table}

\end{itemize}
In \hyperref[B1]{\Block{1}},  we show how to calculate a Laguerre mesh of $N=50$ points together with the corresponding weights using 100-digits arithmetic.


\begin{figure}[h]
	\centering
	\includegraphics[width=0.6\columnwidth]{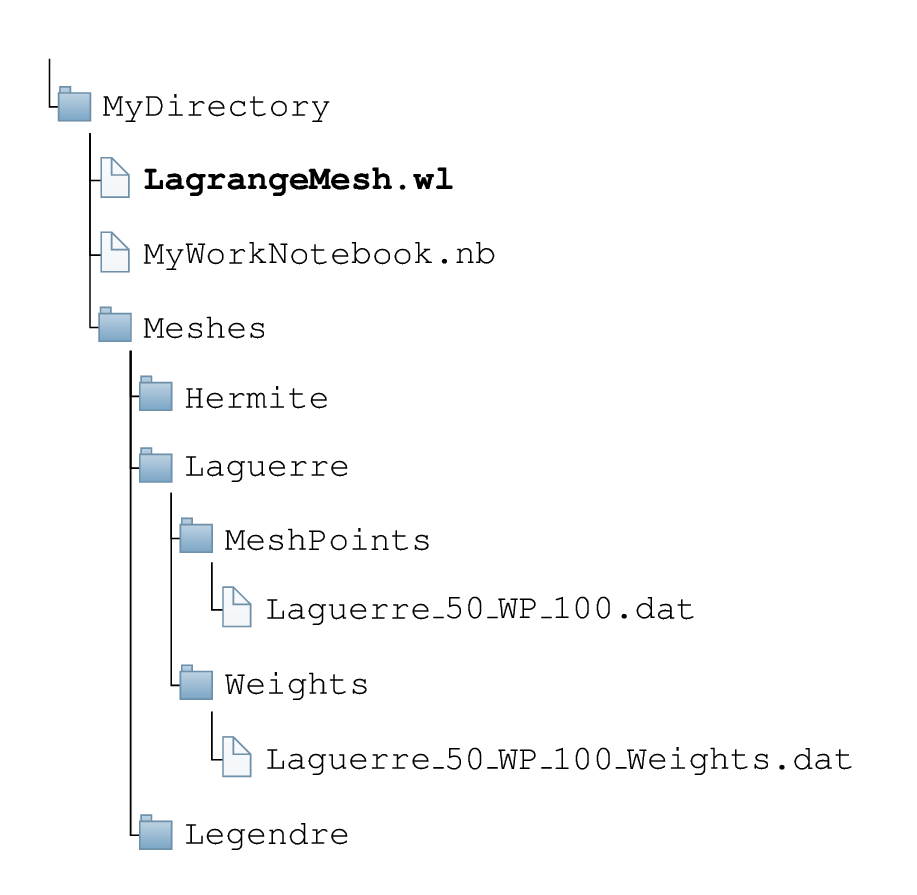}
	\caption{Tree diagram that shows the location of the stored mesh points and weights. The presence of the file \mtext{LagrangeMesh.wl} at the level of \mout{MyWorkNotebook.nb} is optional, it is only required if we are loading the package temporarily, see text. Otherwise, it can be removed from  \mout{MyDirectory}. All directories inside \mout{MyDirectory} will be automatically created on the first use of the command \mcom{BuildMesh}.}
	\label{Fig:location}
\end{figure}

As output, shown in \hyperref[B1]{\Blocko{1}}, the program prints on screen the  name of two files that were generated and stored. They contain the mesh points and weights, respectively. The parameters that characterize the calculation are specified in the  given name. Once created, files will be automatically stored in specific directories, so they can be used in future calculations without calculating them again.  The location of those files is shown in the tree diagram in Fig. \ref{Fig:location}. Using the command \mtext{BuildMesh} once, all necessary directories  will be automatically created.

\subsubsection{\mcom{AvailableMeshQ}}
Once several meshes are constructed and stored, we can generate an ordered table that shows them on screen. The latter is the primary purpose of the command. The secondary is to check if a particular mesh exists: it delivers on screen \mtext{True} in the case it does, and \mtext{False} otherwise. The syntax for this command is the following:
\begin{center}
	\small
	\mcom{AvailableMeshQ[Type,Options]}
\end{center}
\begin{itemize}[label=$\bullet$,leftmargin=*]
	\item	\mcom{Type} specifies the classical polynomial, it can only take the value of the following strings \textcolor{gray}{\mtext{"Legendre"}},  \textcolor{gray}{\mtext{"Laguerre"}}, or   \textcolor{gray}{\mtext{"Hermite"}}. 
	
	\item There are three \mtext{Options} that can be specified for  \mcom{AvailableMeshQ}. They control the output printed on screen. Each option is described in Table \ref{Options:AvailableMesh}.
\end{itemize}

\begin{table}[ht]
	\caption{Options for the command \mcom{AvailableMeshQ}.}
 \resizebox{\textwidth}{!}{	{\begin{tabular}{@{}clc@{}} \toprule
			\mcom{Option}& Description & Default Value  \\ \colrule
			\mcom{Dimension} &If we are interested in displaying meshes of a  particular dimension, & \mcom{False}  \\
			&this  option may be useful. If there is no mesh with such dimension,  \\
			& the output will be  \mtext{False} \\ \\
			\mcom{PrintMesh} &If \mtext{Dimension} and \mtext{WorkingPrecision} are specified, the corresponding   & \mcom{False} \\ 
			&   mesh will be displayed on screen.\\ 
			\\
		\mcom{PrintDomain}  &If \mtext{Dimension} and \mtext{WorkingPrecision} are specified, the  corresponding& \mcom{False} \\ 
			& smallest and greatest mesh points will be displayed on screen.  \\ \\
		\mcom{WorkingPrecision}	 & It specifies to look for meshes with a given number of digits in accuracy.    & \mcom{MachinePrecision} \\ 
			& If only this option and \mtext{Dimension} are specified, the output will be\\ 
			& will be \mtext{True} or \mtext{False}.\\
			\botrule
		\end{tabular}}}
	\label{Options:AvailableMesh}
\end{table}

\begin{figure*}[h]
	\centering
	\includegraphics[scale=0.9]{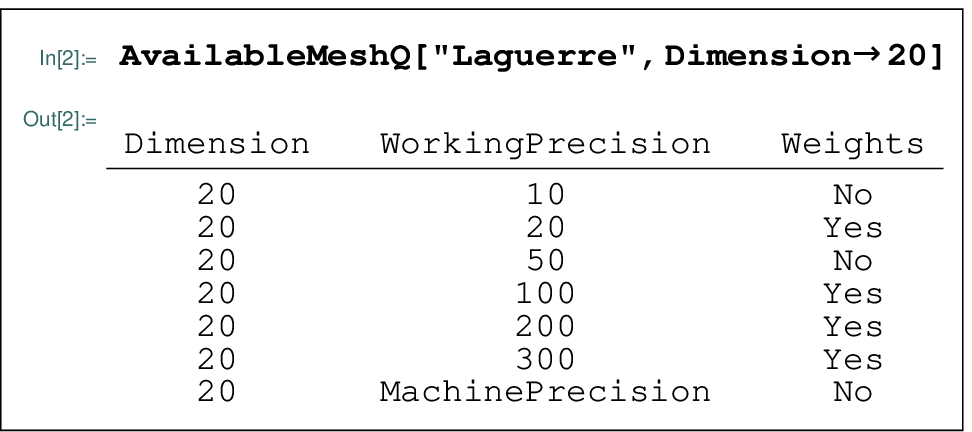}
	\label{B2}
\end{figure*}

As an example, below we present a typical block of code with output \hyperref[B2]{\Blocko{2}} generated by the command \mtext{AvailableMeshQ}.  It consists  of all meshes stored with \mcom{Dimension$\rar$20}.  

\subsubsection{\mcom{LagMeshEigenvalues}}
The main purpose of this command is to calculate the desired number of eigenvalues of the Schr\"odinger equation  (\ref{SEq}) defined in $(a,b)$ for a given potential function $V(x)$. For a given domain, if the corresponding mesh was not previously constructed,  \mcom{LagMeshEigenvalues} will build it automatically by  calling \mcom{BuildMesh} internally. Therefore, the manual usage of \mcom{BuildMesh} can be avoided by the user. After calculations, an ordered list\footnote{In ascending order, starting from the lowest eigenvalue.} with the first approximate eigenvalues will be printed on screen. The syntax of  \mcom{LagMeshEigenvalues} is the following:  
\begin{center}
	\small
	\mcom{LagMeshEigenvalues[V[\textcolor{mygreen}{x}],\{\textcolor{mygreen}{x},a,b\},NLevels,Dimension,Options]}
\end{center}
where
\begin{itemize}[label=$\bullet$,leftmargin=*]
	\item	\mcom{V[\textcolor{mygreen}{x}]} specifies the potential  $V(x)$ of the Schr\"odinger equation (\ref{SEq}). It must be a numerical expression with one degree of freedom, in this case denoted by \mcom{\textcolor{mygreen}{x}}.
	\item \mcom{\{\textcolor{mygreen}{x},a,b\}} defines the domain in the variable \mcom{\textcolor{mygreen}{x}} that appears in the potential. The program automatically will select, construct, and map the appropriate mesh for the given domain. For a finite interval, semi-infinite and infinite, the Legendre, Laguerre, and Hermite are used, respectively. 
	
	\item  \mcom{NLevels} represents the number of the lowest eigenvalues that we desire to calculate. Naturally, its value should be a positive integer number.
	\item  \mcom{Dimension} defines the size of the mesh, i.e., the number of mesh points. 
	
	\item There are seven  \mcom{Options} that can be set by the user. They are presented in Table \ref{Options:eigenvalues} together with a brief description. In addition, all available options of the build-in command \mcom{Eigenvalues} are incorporated into \mcom{LagMeshEigenvalues}. In general terms, they can be useful to reduce CPU times. Details can be found in the Mathematica$^\circledR$ documentation.		
\end{itemize}

There is one important point that we would like to stress. In practical applications, the CPU time and accuracy in approximate eigenvalues are mainly\footnote{The option \mcom{Method} will play an important role in applications, see text.} controlled by three parameters: \mcom{Dimension}, \mcom{WorkingPrecision}, and \mcom{Scaling}. These options make \mcom{LagMeshEigenvalues} a versatile command when the user requires fast but highly accurate calculations. Therefore, choosing \textit{appropriate} values for those options is crucial for the performance of the LMM.  Explicit examples presented in the next Section  will show which considerations must be taken into account when choosing the value of such options.

\begin{table}[ht]
	\caption{Options for  (i) \mcom{LagMeshEigenvalues}, (ii) \mcom{LagMeshEigenfunctions}, and (iii) \mcom{LagMeshEigensystem}. Options marked by $^\dagger$ are only available for commands (ii) and (iii). }
 \resizebox{\textwidth}{!}{	{\begin{tabular}{@{}clc@{}} \toprule
			\mcom{Option}& Description & Default Value  \\ \colrule
			\mcom{CoefficientsOnly}$^\dagger$ &If it is set to \mtext{True}, the output will be the coefficients $c_k$\textcolor{black}{'s} that&\mcom{False} \\
			& solve the Lagrange equations (\ref{lagrange_eqs}).  \\ \\
		\mcom{DiscreteFunction}$^\dagger$&If it is set to \mtext{True}, the output will be a list that contains a discrete  & \mcom{False} \\ 
			&version of the wavefunctions according to 
			(\ref{discrete}).    \\ 
			\\
			\mcom{ExpectationValue}$^\dagger$ &Function for which the expectation value is calculated via  the& \mcom{False} \\ 
			&  approximation (\ref{expectation}).   \\ \\
				\mcom{Mass}	 & It fixes the value of the mass $(m)$, see (\ref{SH}). Its value \textcolor{black}{may} be    & \mcom{1} \\ 
			&a \textcolor{black}{complex} number.   \\  \\
					\mcom{PotentialShift} & Shifts the potential $V(x)$ by a real constant. Final results for    & \mcom{0} \\ 
			&the spectrum do not depend on the value of this option, see text.  \\  \\
				\mcom{Scaling}	 &  Its value corresponds to the positive scaling parameter $h$, defined & \mcom{1} \\ 
			& in (\ref{scaling}).  \\ 	 \\
				\mcom{WorkingPrecision}	 &It specifies how many digits of precision should be maintained in    & \mcom{MachinePrecision} \\ 
			&internal computations. Thus, it controls the numeric accuracy  \\ 
			&of the approximate spectrum. \\
			\botrule
	\end{tabular}}}
	\label{Options:eigenvalues}
\end{table}

\subsubsection{\mcom{LagMeshEigenfunctions}}
This command delivers as output a list that contains the lowest approximate eigenfunctions of the Schr\"odinger equation  (\ref{SEq}) defined in $(a,b)$ for a given potential function $V(x)$. By default, they are presented normalized and in the form (\ref{Continuos}).  If the corresponding mesh used for calculations was not previously constructed,  \mcom{LagMeshEigenfunctions} will build it automatically calling \mcom{BuildMesh} internally. Therefore, the manual usage of \mcom{BuildMesh} can be avoided by the user. After calculations, an ordered list\footnote{In ascending order, starting from the eigenfunction associated with the  lowest eigenvalue.}  with the first approximate eigenfunctions will be printed on screen. The syntax of the command is the following:  

\begin{center}
	\small
	\mcom{LagMeshEigenfunctions[V[\textcolor{mygreen}{x}],\{\textcolor{mygreen}{x},a,b\},NLevels,Dimension,Options]}
\end{center}

The description of each element written  above was already presented  above when discussing the usage of  \mcom{LagMeshEigenvalues}. Therefore, the reader is referred to that Section for discussion. The available Options for this command are shown in Table \ref{Options:eigenvalues}.  In addition, all available options of the built-in command \mcom{Eigenvectors} are incorporated into \mcom{LagMeshEigenfunctions}.

As a general \textit{heuristic} remark, we stress the following.  \textit{Appropriate}\footnote{For example, they lead to accurate eigenvalues.} values for the options  \mcom{WorkingPrecision}, \mcom{Dimension}, and \mcom{Scaling} used in \mcom{LagMeshEigenvalues}, are frequently suitable  for the command \mcom{LagMeshEigenfunctions}. In this way, the calculation of   eigenvalues serves as guidance to find accurate eigenfunctions. This idea goes both ways: appropriate values for \mcom{LagMeshEigenfunctions} are also appropriate for \mcom{LagMeshEigenvalues}.

\subsubsection{\mcom{LagMeshEigensystem}}
This command delivers simultaneously the desired number of eigenvalues and eigenfunctions of the Schr\"odinger equation  (\ref{SEq}) defined in $(a,b)$ for a given potential function $V(x)$. The syntax  is  the same as for the command \mcom{LagMeshEigenfunctions},  namely
\begin{center}
	\small
	\mcom{LagMeshEigensystem[V[\textcolor{mygreen}{x}],\{\textcolor{mygreen}{x},a,b\},NLevels,Dimension,Options]}
\end{center}
In general terms, \mtext{LagMeshEigensystem}  can be regarded as the combination of the previous two. Therefore, it contains all available options for  \mcom{LagMeshEigenvalues} and \mcom{LagMeshEigenfunctions}, see Table \ref{Options:eigenvalues}.  In addition, all available options of the built-in command \mcom{Eigensystem} are incorporated into \mcom{LagMeshEigensystem}.

\subsection{Worked Examples}
\label{WorkedExamples}
In this Section, we present different applications of the package for some well-known and relevant systems. The following examples will illustrate and explore how the values of the options can be crucial to obtain  accurate eigenvalues and eigenfunctions. A comparison with results from the  literature is made for some examples.  Finally, we encourage the reader to check the supplementary material (\mcom{WorkedExamples.nb}) to reproduce some of the following calculations. Throughout the rest of this work,  unless  we specify otherwise, the mass is set to $m=1$.
\subsubsection{Exactly Solvable  Potentials}
We present approximate  eigenvalues and eigenfunctions obtained by the \mtext{LagrangeMesh} package  and compare them with the exact spectrum. We have chosen  three potentials that are representative of each possible domain: finite, semi-infinite, and infinite.  The following examples are helpful to become familiar with the commands and basic options of the package. 

\paragraph{Particle in a Box.}
Let us begin with the simplest one-dimensional potential that holds infinitely many bound states: a particle confined to a rigid box of length $L$. This potential is described by  
\begin{equation}
	V(x)\ =\ \begin{cases}           0,\quad &0\leq|x|\leq L\\
		\infty,\quad &\text{otherwise}
	\end{cases}
\end{equation}	
The exact solution of the Schr\"odinger equation  leads to the following expression for the eigenvalues
\begin{equation}
	E_n\ =\ \frac{\pi^2n^2}{2L^2}\ ,\qquad n=1,2,...\ .
	\label{Infinitewell}
\end{equation}
On the other hand, using  \mcom{LagMeshEigenvalues} let us calculate the first 3 lowest eigenvalues using $N=50$ mesh points and an arithmetic of 20 digits. To do so, we take $L=1$ without loss of generality. In this case, the syntax looks like

\begin{figure*}[h]
	\centering
	\includegraphics[scale=0.9]{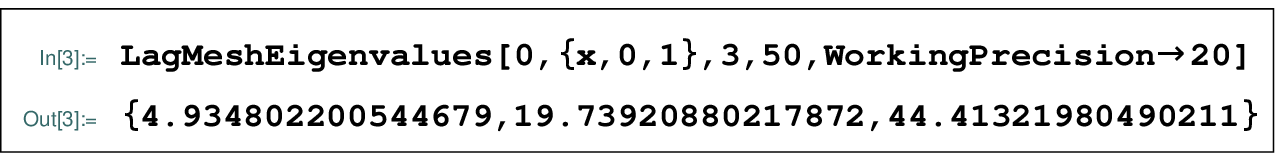}
	\label{B3}
\end{figure*}

When comparing with the exact spectrum (\ref{Infinitewell}), one will note that  \textcolor{black}{the deviation of the} approximate eigenvalues is of order $10^{-15}$. In this case, the error can be reduced (at least) up to $10^{-90}$ by increasing  \mtext{WorkingPrecision$\rar$200}, but keeping the number  of mesh points fixed to $N=50$. 

\begin{figure}[h]
	\centering
	\includegraphics[scale=0.9]{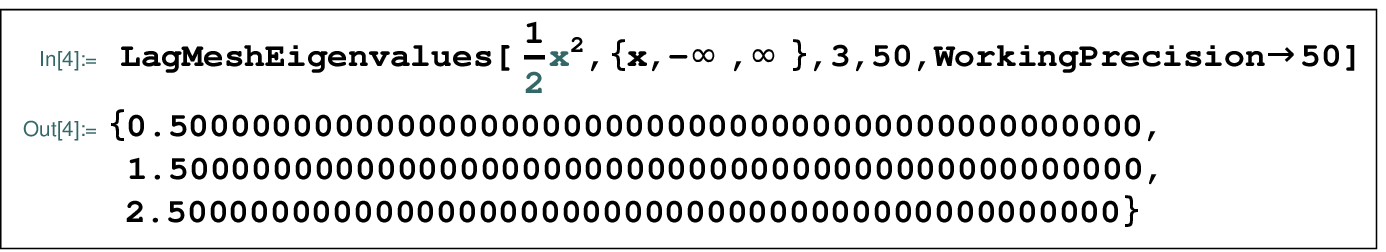}
	\label{B4}
\end{figure}

\paragraph{Harmonic Oscillator.}
This elementary system is described by the potential
\begin{equation}
	V(x)\ =\ \frac{1}{2}x^2\ ,\qquad -\infty<x<\infty\ .
	\label{harmonic}
\end{equation}
The corresponding Schr\"odinger equation is exactly solvable, which means that the spectrum can be found exactly. In particular, eigenvalues and (normalized) eigenfunctions are 
\begin{equation}
	E_n\ =\ n\ +\ \frac{1}{2}\ ,\qquad\psi_n(x)\ =\ \frac{1}{\pi^{1/4}\sqrt{2^nn!}}H_n(x)e^{-x^2/2} \qquad n=0,1,2,...\ ,
	\label{exactHO}
\end{equation}
where $H_n(x)$ is the $n$-th Hermite polynomial.
First, let us compute the first three  eigenvalues using the  input \hyperref[B4]{\Block{4}}. 

We can immediately see the excellent agreement between the formula (\ref{exactHO}) and the output of $\mcom{LagMeshEigenvalues}$: it is  consistent with the requested accuracy.  Although we used $\mcom{WorkingPrecision}\rar\mtext{50}$, the output  \hyperref[B4]{\Blocko{4}} gives energies with 47-48 exact digits. Therefore, some numerical errors were accumulated during internal computations. Once again, they can be reduced by increasing \mtext{WorkingPrecision}.

\begin{figure*}[h]
	\centering
	\includegraphics[scale=0.9]{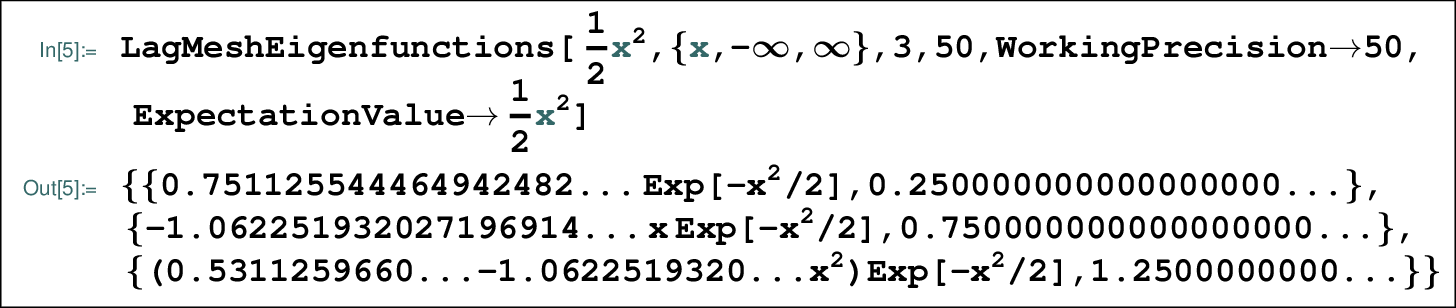}
	\label{B5}
\end{figure*}

Now, we focus on calculating the approximate first three eigenfunctions and the expectation value of the potential (\ref{harmonic}). To do so, we have input \hyperref[B5]{\Block{5}}.

Output \hyperref[B5]{\Blocko{5}} of the program delivers  a list with normalized eigenfunctions   and the expectation \textcolor{black}{values} of $V(x)$. Note that we have dropped digits in order to  simplify the presentation. 
For the  harmonic oscillator, the LMM leads to the exact eigenvalues and eigenfunctions  within the desired accuracy\footnote{For further details, see \textcolor{black}{Ref.} \cite{Baye2015}.}. This is not a surprise, looking at  (\ref{hermite}), it is clear that the first $N-1$ eigenfunctions of the harmonic oscillator can be written as a linear combination of  Lagrange functions $\{f_i(x)\}_{i=1}^{N}$. The expectation values\footnote{The expectation value of the potential results in $E_n/2$ according to the Virial theorem, see Ref. \cite{Sakurai}. } of $V(x)$ are accurate with the same accuracy obtained for the energy: 47-48 exact digits.

\paragraph{Hydrogen Atom.}

As mentioned above,  the (reduced) 3-dimensional radial Schrödinger equation can be solved using  a Laguerre mesh and the corresponding Lagrange functions (\ref{Laguerre}).  Let us consider the Coulomb potential, 
\begin{equation}
	V(r)\ =\ -\frac{1}{r}\ ,\qquad 0\leq r<\infty\ .
	\label{Coulomb}
\end{equation}
In spherical coordinates, it is well known that eigenfunctions can be labeled by three quantum numbers $(n,\ell,m)$. However, the eigenvalues  only depend  on the principal quantum number $n$, namely
\begin{equation}
	E_n\ =\ -\frac{1}{2n^2}\ ,\qquad n=0,1,2,...\ .
\end{equation}
Using input \hyperref[B4]{\Block{6}}, we calculate the first six eigenvalues with angular momentum $\ell=0$. \textcolor{black}{Note that $\ell$ will be denoted by $\mcom{l}$ inside the blocks of code.} For completeness, we have introduced the centrifugal potential despite the fact it vanishes. 

\begin{figure*}[h]
	\centering
	\includegraphics[scale=0.9]{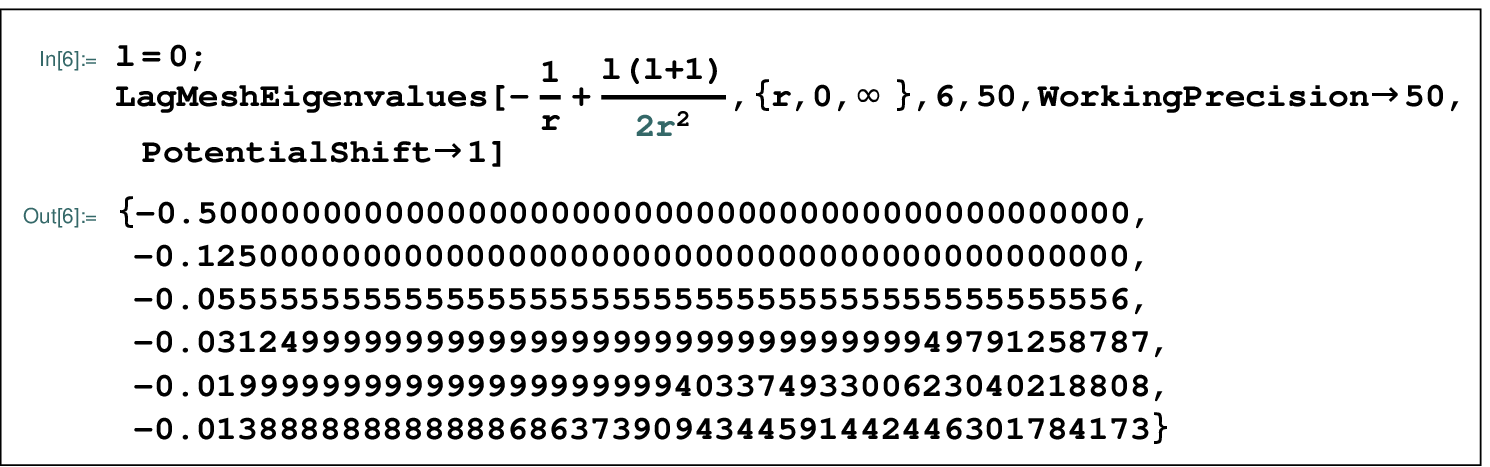}
	\label{B6}
\end{figure*}

There is one important remark in the code shown above.  Only for \textit{unbounded from below} potentials \textcolor{black}{- for example, (\ref{Coulomb}) -}  defined in $[a,b]$, it is mandatory to use a non-zero value of the option \mcom{PotentialShift} to obtain correct results. This value should be chosen in such a way that the shifted potential
\begin{equation}
	V(x)\ +\ \mcom{PS}\ ,\qquad \mcom{PS} =\mcom{OptionValue[PotentialShift]}\ ,
\end{equation}
has a positive spectrum.  In the present case, any value of \mcom{PS}$\,>1/2$  does the work. This is  merely a technical issue related to  Mathematica$^\circledR$ and not with the LMM. Naturally,  final results are independent of the value of \mcom{PotentialShift}.

Within the requested accuracy, we note in output \hyperref[B6]{\Block{6}} an excellent agreement between the exact energies of the first three levels and those predicted by the LMM. Certainly, the remaining three eigenvalues are close to the exact ones,  but numerical errors are evident. In this case, increasing the value of \mcom{WorkingPrecision} is not the solution. However, there are two ways to reduce the errors in this case: (i) increase the dimension of the mesh considering $N>50$; (ii) choose an appropriate value for \mtext{Scaling}. We will postpone the discussion of these two approaches to the following Section.

\subsubsection{Some Non-Solvable Potentials}
In this Section, we consider some representative non-solvable potentials and show how to tackle them with the \mtext{LagrangeMesh} package. In what follows, we explore the different \mtext{Options} and show how some of them play a fundamental role to obtain highly accurate results for either eigenvalues or eigenfunctions, despite the fact that exact solutions are unknown. 

\paragraph{Quartic Anharmonic Oscillator.}

The first example we consider is one of the most studied systems in quantum mechanics: the celebrated quartic anharmonic potential\footnote{The present discussion is based on Ref. \cite{Turbiner2021}.}
\begin{equation}
	V(x)\ =\ \frac{1}{2}x^2\ +\ \frac{1}{4}x^4\ ,\qquad -\infty<x<\infty\ .
\end{equation}
The discussion is focused on the lowest eigenvalue: the ground state energy denoted by $E_0$. We study the accuracy of the method as a function of the number of mesh points $N$ keeping fixed  \mcom{WorkingPrecision}$\rar$\mtext{300}. As we will see,  when considering  a sufficiently large value for  \mcom{WorkingPrecision}, we can ensure that our results are not \textit{contaminated} by a loss of accuracy due to internal  arithmetic manipulations. Then, the convergence of the approximate $E_0$ will occur as $N\rar\infty$.

We take some representative dimensions from $N=25$ to \textcolor{black}{$N=2020$.} 
For the smallest mesh $N=25$, the block of code looks like

\begin{figure*}[h]
	\centering
	\includegraphics[scale=0.9]{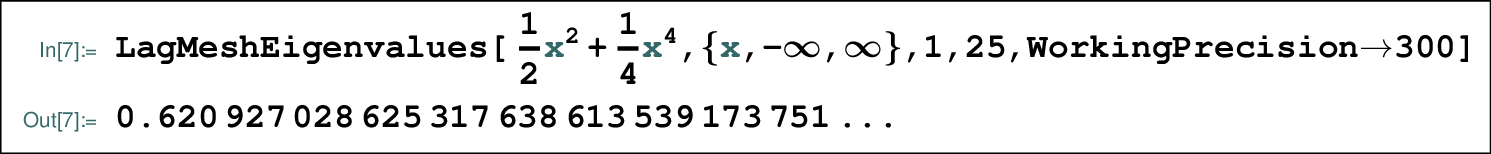}
	\label{B7}
\end{figure*}

We have dropped a considerable number of decimal digits in the output \hyperref[B7]{\Blocko{7}} to simplify the presentation. The modification of the previous block for considering  larger meshes is trivial. For  a large number of mesh points, naturally calculations require more CPU time. In fact,  the most time-demanding calculation corresponds to the mesh points.\footnote{Assuming that all meshes are already determined, the CPU time needed by the LMM scales as $N^3$. } For example, calculating the largest Hermite mesh with $N=2000$ with accuracy  \mcom{WorkingPrecision}$\rar$\mtext{300} takes one whole day \textcolor{black}{using a nowadays standard computer}. Meanwhile, the realization of the LMM only takes 40 minutes. More details about CPU times can be found in Ref. \cite{Turbiner2021}. Results for all  considered meshes are shown in (\ref{QAHO}). There, we have used the following notation,
\begin{equation}
	\underset{N_Z}{X}:\quad X= \text{Digit}\ ,\quad N=\text{Number of mesh points } \ ,\quad Z = \text{Decimal place of }X\ .
\end{equation}	
Digit $X$ indicates the maximal digit in energy which is reproduced with a given number of mesh points $N$. For example, with $N=50$ the ground state energy is obtained with 14 exact decimal digits. The maximal accuracy is reached with $N=2000$ mesh points \textcolor{black}{as it was} confirmed with $N=2020$. 
\begin{align}
	E_0\ =\  \textcolor{gray}{0.}&
	\textcolor{gray}{620\,927\,0
		\textcolor{black}{\underset{\mathclap{{25{{}_{8}}}}}{\textBF{2}}}
		9
		\,825\,7
		\textcolor{black}{\underset{\mathclap{{50{{}_{14}}}}}{\textBF{4}}}
		8\,660\,8
		\textcolor{black}{\underset{\mathclap{{75{{}_{20}}}}}{\textBF{5}}}
		8\,035\,
		\textcolor{black}{\underset{\mathclap{{100{{}_{25}}}}}{\textBF{7}}}
		32\,987\,12
		\textcolor{black}{\underset{\mathclap{{150{{}_{33}}}}}{\textBF{0}}}
		\,698\,200\,
		017\,2
		\textcolor{black}{\underset{\mathclap{{200{{}_{44}}}}}{\textBF{5}}}
		3\,619}\nonumber \\
	&\textcolor{gray}{1
		\textcolor{black}{\underset{\mathclap{{250{{}_{50}}}}}{\textBF{3}}}
		8\,982\,542\,3
		\textcolor{black}{\underset{\mathclap{{300{{}_{59}}}}}{\textBF{6}}}
		7\,325\,062\,962\,74
		\textcolor{black}{\underset{\mathclap{{400{{}_{72}}}}}{\textBF{8}}}
		\,188\,768\,883\,979\,39
		\textcolor{black}{\underset{\mathclap{{500{{}_{87}}}}}{\textBF{1}}}
		\,351\,303\,479}\nonumber\\
	&\textcolor{gray}{456\,083\,601\,618\,760
		\,073\,476\,624\,891\,085\,768\,308\,099\,065\,938\,402}\nonumber\\
	&	\textcolor{gray}{
		\textcolor{black}{\underset{\mathclap{{1000{{}_{145}}}}}{\textBF{5}}}
		80\,084\,530\,397\,024\,737\,474\,347\,663\,406\,954\,493\,075\,566\,093\,052}\nonumber\\
	&	\textcolor{gray}{396\,859\,302\,472\,486\,392\,601\,975\,136\,357\,293\,108\,871\,529\,43
		\textcolor{black}{\underset{\mathclap{{1900{{}_{237}}}}}{\textBF{9}}}
		\,117}\nonumber\\
	&\textcolor{gray}{092\,27\textcolor{black}{\underset{\mathclap{{2000_{{}_{246}}}}}{\textBF{5}}}}\ .\nonumber\\
	\label{QAHO}
\end{align}

In this case, the rate of convergence is about 10-11 correct digits with respect to an increment of the number of mesh points in 100. Smaller meshes can reach the maximal accuracy if an appropriate value of \mtext{Scaling} is used. Consequently, CPU times are reduced. We will discuss this aspect based on another example, see below.

\begin{figure*}[h]
	\centering
	\includegraphics[scale=0.9]{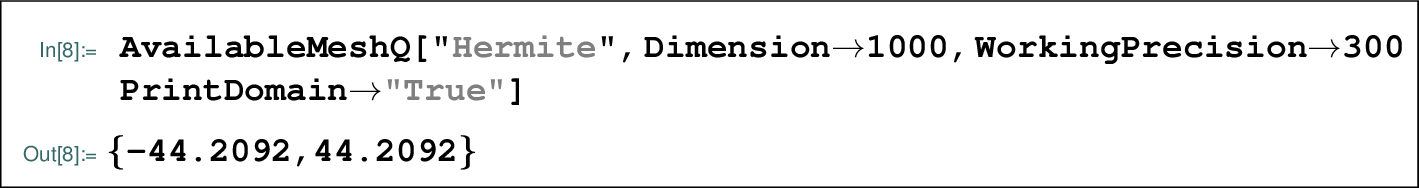}
	\label{B8}
\end{figure*}

\paragraph{Quartic Double Well Potential.}
Now we consider the potential
\begin{equation}
	V(x)\ =\ \frac{1}{2}x^2(1-gx)^2\ ,\qquad -\infty<x<\infty\ ,
	\label{DW}
\end{equation}
which  is a double well  with  two degenerate minima  located at $x=0$ and $x=1/g$. When $g\neq0$, 
it is well-known that there is  an exponentially small separation between the energies of the ground \textcolor{black}{($E^+$)} and first excited ($E^-$) states \cite{Landau}. Let us define and denote this energy gap as
\begin{equation}
	\Delta E\ =\ E^+-E^-\ .
\end{equation}
By means of semi-classical analysis, \textcolor{black}{see Ref. \cite{Zinn}}, we know the first terms of the asymptotic resurgent expansion of $\Delta E$, namely
\textcolor{black}{
\begin{equation}
	\Delta E_{SC}\ =\  \frac{2}{\sqrt{\pi}g}e^{-\frac{1}{6g^2}}\left(1\ - \frac{71}{12}g^2\ +\ \mathcal{O}(g^2)\right)\ +\ \mathcal{O}\left(e^{-\frac{1}{3g^2}}\right)\ ,
	\label{gap}
\end{equation}}
where the sub-index $SC$ stands for semi-classical. Let us investigate if the LMM can capture the exponentially small contribution in the energy gap for $g=1/30$. Let us point out that the degenerate minima are $x=0$ and $x=30$. Therefore, the mesh that we need to consider must cover these two points. It turns out that $N=1000$ is appropriate as shown in the following input where we print on screen the smaller and largest mesh points, see \hyperref[B8]{\Block{8}}  and \hyperref[B8]{\Blocko{8}}.

Now we calculate the first two eigenvalues using the Hermite mesh, see  \hyperref[B9]{\Block{9}} below.

\begin{figure*}[h]
	\centering
	\includegraphics[scale=0.9]{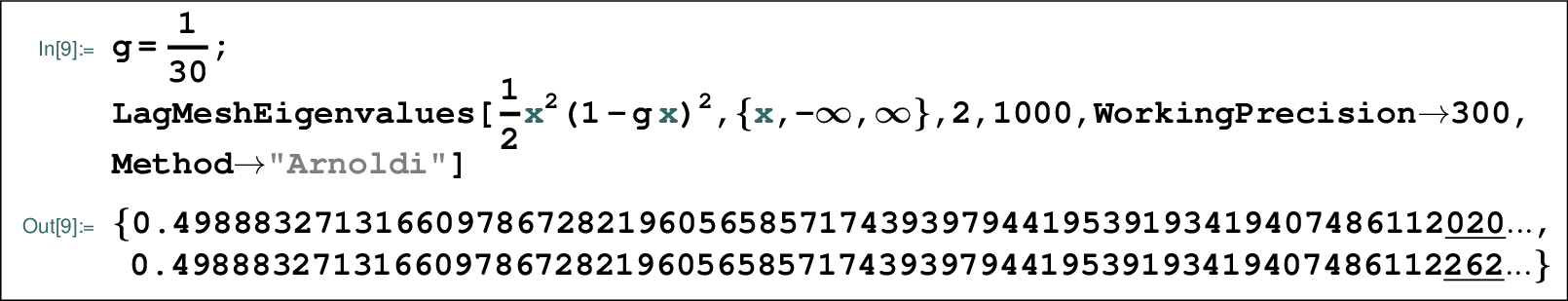}
	\label{B9}
\end{figure*}

We have introduced the built-in option \mcom{Method$\rar$\textcolor{gray}{"Arnoldi"}} to reduce CPU times: from 1.2 hrs  to 5 min. In addition,  
the last digits of the output were removed. From the  \textcolor{black}{first two}  eigenvalues, we can estimate the energy gap and compare with (\ref{gap}):
\begin{equation}
	\Delta E_{LMM}\ = \ 2.4129\times10^{-64}\ , \qquad \Delta E_{SC}\ =\ 2.4128\times10^{-64}\ .
\end{equation}
We can see that there is an excellent agreement between both estimates\footnote{SC-estimate was calculated neglecting terms of $\mathcal{O}(g^2)$ and $\mathcal{O}(e^{-1/3g^2})$ in (\ref{gap}).}. Let us now investigate how the wavefunctions look like with respect to $x$.  Since the dimension of the basis is quite large $(N=1000)$, it is convenient to use the discrete version of the wavefunction to avoid loss of accuracy. To do so, we use input \hyperref[B10]{\Block{10}} shown below.

\begin{figure*}[h]
	\centering
	\includegraphics[scale=0.9]{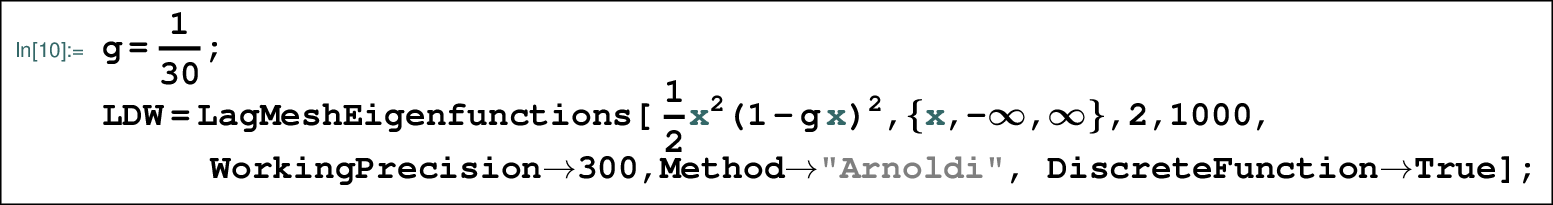}
	\label{B10}
\end{figure*}

In the previous block of code, we have stored the discretized wavefunctions in a list called \mtext{LDW}. Then, the plots of the first two wavefunctions can be easily obtained using the built-in command \mtext{ListPlot},

\begin{figure*}[h]
	\centering
	\includegraphics[scale=0.9]{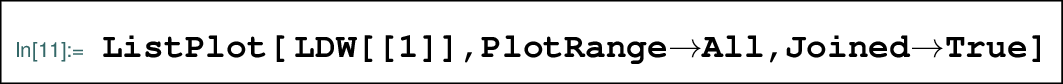}
	\label{B11}
\end{figure*}

\begin{figure}[h]
	\centering
	\includegraphics[width=0.65\columnwidth]{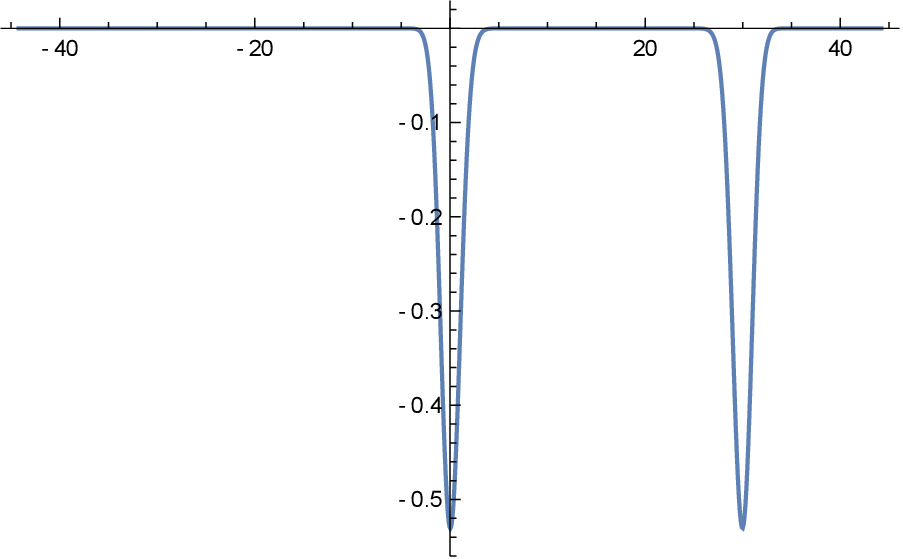}
	\caption{Approximate  wavefunction of the ground state in  the double-well  potential (\ref{DW}). Plot generated by \hyperref[B10]{\Block{10}} and \hyperref[B11]{\Block{11}}.}
	\label{Fig:Even}
\end{figure}

\begin{figure}[h]
	\centering
	\includegraphics[width=0.65\columnwidth]{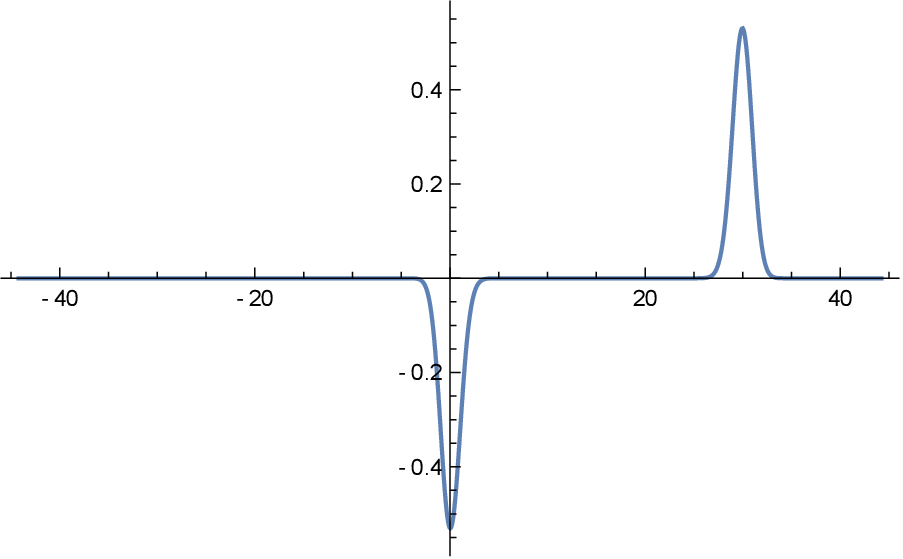}
	\caption{Approximate wavefunction of the  first excited state in the double-well  potential (\ref{DW}). Plot generated via \hyperref[B10]{\Block{10}} and minimal modifications of  \hyperref[B11]{\Block{11}}. }
	\label{Fig:Odd}
\end{figure}

The output   of the previous block \hyperref[B11]{\Block{11}} corresponds to the plot of the ground state eigenfunction shown in Fig. \ref{Fig:Even}. With minimal modifications, the same block can be used to plot the first excited wavefunction, see Fig. \ref{Fig:Odd}. 

There is one point that should be emphasized about the Hermite mesh used in this example. By construction, it always will be symmetrical with respect to the vertical axis ($x=0$). As a consequence, the discrete version of the  wavefunction will
be defined for mesh points inside \textcolor{black}{the domain} $-x_N\leq x\leq x_N$. However, it can be seen from Figs. \ref{Fig:Even} and \ref{Fig:Odd} that the \textcolor{black}{wavefunctions} do not share the same symmetry of the domain. Furthermore, wavefunctions are \textcolor{black}{negligible} for $x<-10$. To improve the performance of the LMM, it is natural to shift/translate the potential $V(x)\rar V(x+1/2g)$ to make it symmetrical under $x\rar -x$. Naturally,  final results for the spectrum do not depend on the shift. For instance, $N=500$ can provide the same accuracy for the energy gap in comparison with previous calculations with $N=1000$. In this case, CPU time was reduced  from 20 min to 50 s. 


\paragraph{Shell-Confined Hydrogen Atom.}

\begin{figure*}[h]
	\centering
	\includegraphics[scale=0.9]{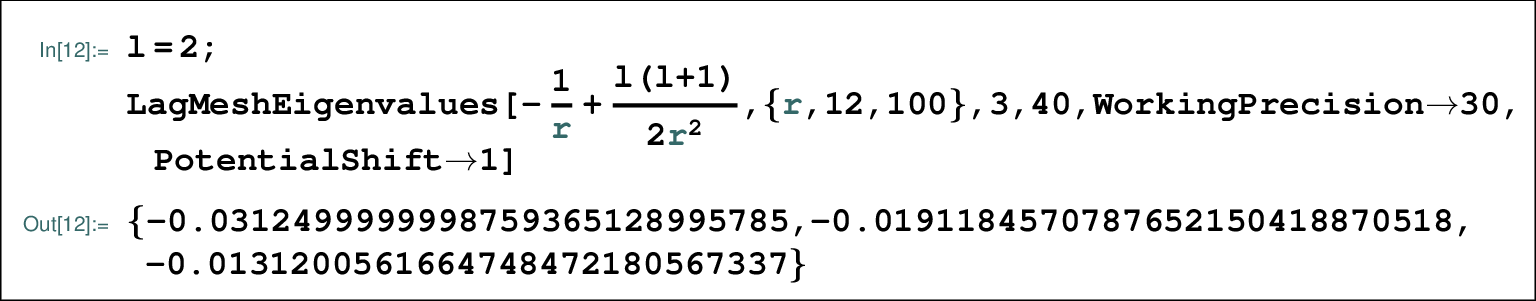}
	\label{B12}
\end{figure*}

This system is described by the Schr\"odinger equation (\ref{SEq}) with the usual Coulomb potential  $V(r)=-1/r$. \textcolor{black}{We assume boundary conditions that} guarantee the shell-type configuration:
\begin{equation}
	\psi(r_1)\ =\ \psi(r_2)\ =\ 0\ ,\qquad 0\leq r_1\ <\ r_2\ .
\end{equation}
Let us take as an example the particular core-shell configuration 
\begin{equation}
	r_1\ =\ 12 \ , \qquad r_2\ =\ 100\ ,
\end{equation}
and calculate the first three energies for angular momentum $\ell=2$. To do so, we use \hyperref[B12]{\Block{12}}.
We confirmed  the results  provided in Ref. \cite{Sen}. We note that the first eigenvalue in the list  is very similar to the energy of the familiar  states\footnote{Associated with  the potential (\ref{Coulomb}).} of the \textit{free} hydrogen atom with principal quantum number $n=4$, which is $E_{4} =-1/32$. This is not a coincidence, the usual wavefunction for the state $4d$ has a  node at $r_1=12$ and another one at $r_2=\infty$. In this example $r_2=100$ \textcolor{black}{mimics} "$\infty$". This phenomenon will be discussed below.

\paragraph{$\mathcal{PT}$-symmetric \textcolor{black}{Cubic Oscillator}.}
For some systems defined on a semi-infinite or infinite interval, it is enough to confine them in a sufficiently large box without obtaining essential changes in the spectrum. We explore such situation based on a  particular example. Let us take,
\begin{equation}
	V(x)\ =\  i\,x^3\ ,\qquad -\infty\  <\ x\ <\ \infty\ .
	\label{PT}
\end{equation}
This potential is invariant under the simultaneous transformations\footnote{$\mathcal{PT}$-symmetry stands for $\mathcal{P}$arity and $\mathcal{T}$ime reversal.} $x\rar-x$ and $i\rar-i$ as well as the kinetic term in the Hamiltonian. It is well-known that the spectrum associated with this potential is positive and discrete, see Ref.  \cite{Bender}. Using the package, we can easily obtain the first eigenvalues with unprecedented accuracy. For this purpose, we use the following block of code.

\begin{figure*}[h]
	\centering
	\includegraphics[scale=0.9]{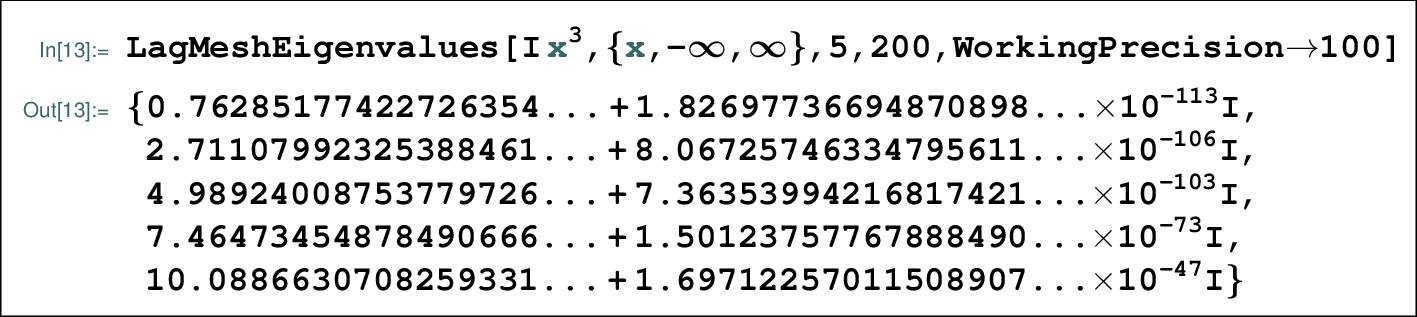}
	\label{B13}
\end{figure*}
Eigenvalues in \hyperref[B13]{\Block{13}} are \textit{essentially} real, with a  small imaginary part. In particular, the first three are consistent with the requested accuracy of 100 exact digits in arithmetic \footnote{Small values of \mcom{WorkingPrecision} can lead to totally wrong results.}. 
Since the potential (\ref{PT}) is defined in the whole real line, it was natural to specify the domain   \mtext{\{\textcolor{mygreen}{x},-\bm{$\infty$},\bm{$\infty$}\}} inside \mcom{LagMeshEigenvalues}. However, with a sufficiently large \textcolor{black}{but} finite domain\footnote{A similar phenomenon occurred in the previous Section for the shell-confined hydrogen atom.}, it is possible to reproduce exactly those results as shown below.

\begin{figure*}[h]
	\centering
	\includegraphics[scale=0.9]{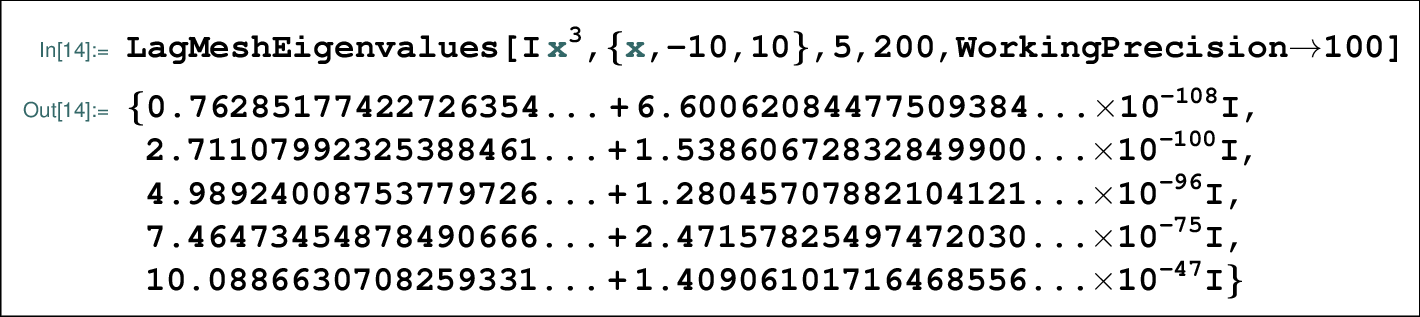}
	\label{B14}
\end{figure*}
The real part of numerical eigenvalues in \hyperref[B14]{\Blocko{14}} is confirmed, meanwhile the imaginary part is different but still small.  Previous results obtained by the LMM overcome the best benchmarks  found in literature, see Ref. \cite{Bender}.

\paragraph{Quasi-Exactly Solvable Double Well.} From Ref. \cite{TurbinerQES}, we extract the one-dimensional potential\footnote{In Ref. \cite{TurbinerQES}, equation (2.1), $a=b=1$, $k=0$, and $n=4$.}:
\begin{equation}
	V(x)\ =\ x^6+2 x^4-18 x^2-1\ ,\qquad \textcolor{black}{-}\infty\ <\ x\ <\ \infty\ .
	\label{QES}
\end{equation}
Degenerate minima are located approximately at $x=\pm1.368$.
For this potential, only the five lowest  eigenfunctions\footnote{In the Hamiltonian operator (\ref{SH}), the mass must be taken as  $m=1/2$.} of even parity can be found in exact form. They can be written as 
\begin{equation}
	\psi_n(x)\ =\ P_n(x^2)e^{-\frac{1}{2}x^2-\frac{1}{4}x^4}\ , \qquad n=0,1,...,4\ ,
\end{equation}
where $P_n(x)$ is a real  polynomial of degree $n$.
Correspondingly, eigenvalues and eigenfunctions can be computed via simple algebra. In particular, eigenvalues read
\begin{gather}
	E_0\ =\ -14.044\,499\,331\ ,\qquad E_1\ =\ \textcolor{black}{-}3.247\,407\,444\ ,\qquad E_2\ =\ 4.623\,648\,530\ ,  \nonumber\\
	E_3\ =\ 17.853\,104\,881\ ,\qquad	E_4\ =\ 34.815\,153\,363\ . 
	\label{eq:c}
\end{gather}
In this example, we now study the effect of  \mcom{Scaling} on the accuracy of the method keeping the number of mesh points fixed $(N=30)$.  First, let us calculate the first nine eigenvalues using \mcom{LagMeshEigenvalues}, and then extract those which correspond to even parity. The input \hyperref[B15]{\Block{15}}, shown below, works for this purpose. 

\begin{figure*}[h]
	\centering
	\includegraphics[scale=0.9]{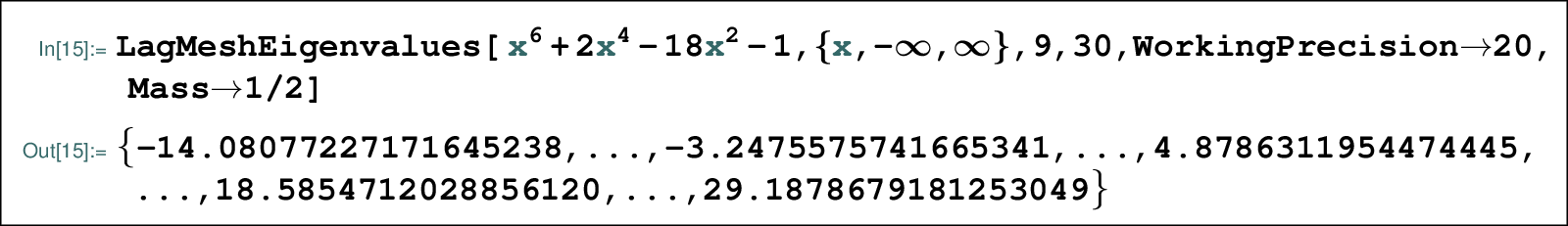}
	\label{B15}
\end{figure*}

\begin{figure*}[h]
	\centering
	\includegraphics[scale=0.9]{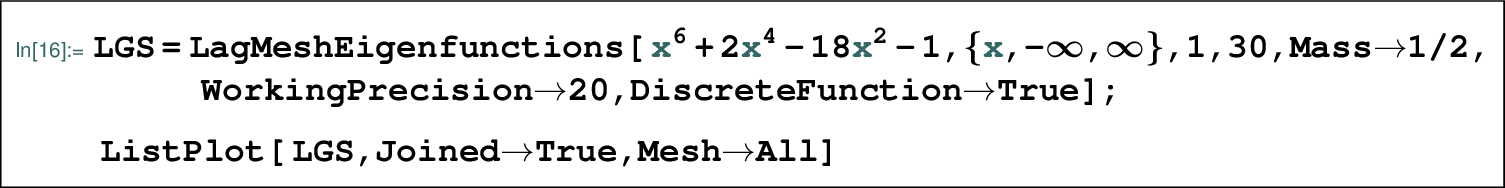}
	\label{B16}
\end{figure*}

Since we are interested in even parity states, we have replaced the energies of odd states with dots in the  output \hyperref[B15]{\Blocko{15}}. Compared with (\ref{eq:c}), we note that numerical results are not very accurate, especially for excited states. The performance of the method can be improved if we choose an adequate scaling parameter. Keeping this target in mind, it is sufficient to calculate and plot the discrete version of the ground state wavefunction. For this purpose, we use \hyperref[B16]{\Block{16}}.

\begin{figure}[h]
	\centering
	\includegraphics[width=0.6\columnwidth]{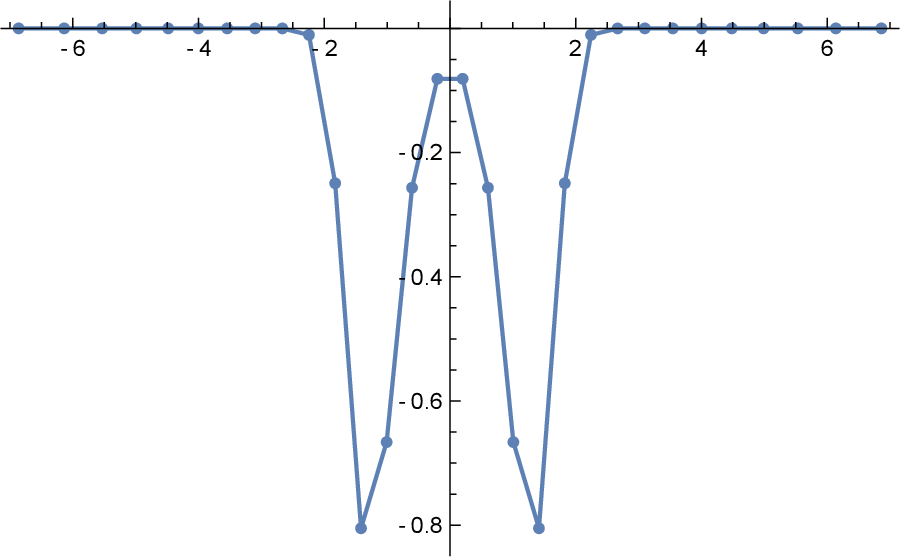}
	\caption{Approximate discrete ground state wavefunction of the quasi-exactly solvable potential (\ref{QES}). For $N=30$, mesh points are distributed in the domain $-7 < x < 7$\ . Plot generated via \hyperref[B16]{\Block{16}}.}
	\label{Fig:DW}
\end{figure}

\begin{figure*}[h]
	\centering
	\includegraphics[scale=0.9]{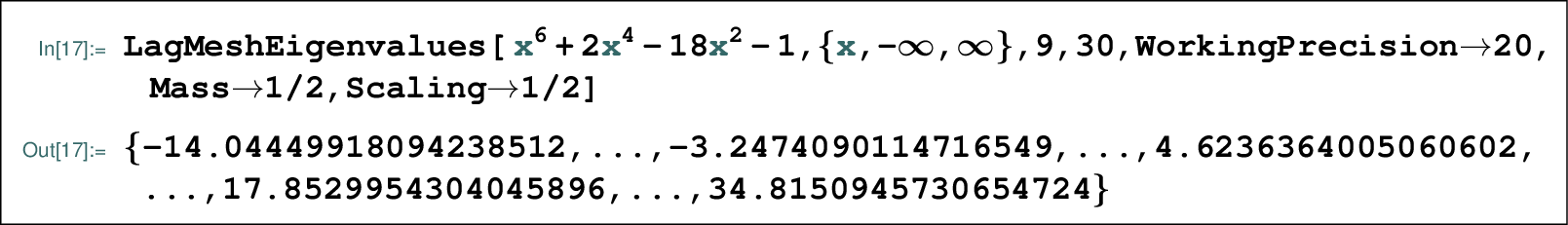}
	\label{B17}
\end{figure*}

The corresponding output is the  plot is presented in Fig. \ref{Fig:DW}.
From this plot, we note that the wavefunction is small  ($\sim10^{-4}$) if $|x|\geq3.5$. Therefore, it is convenient to scale the  mesh to move all mesh points to the region $|x|\leq3.5$, where the function is not too small. This can be done by means of \mcom{Scaling}. For this particular case, $\mcom{Scaling}\rar1/2$ seems to be suitable. Recalculating the spectrum with this scaling, we obtain  more accurate approximate energies, especially for excited states. This can be \textcolor{black}{checked} in the  block of code for \hyperref[B17]{\Block{17}}. 

\paragraph{Rydberg Atoms.} As final example, we show how \mtext{LagMeshEigensystem} works. Let us consider the following radial potential
\begin{equation}
	V(r)\ =\ -\frac{Z_\ell(r)}{r}\ -\ \frac{a_c}{2r^4} \left[1-e^{-(r/r_c)^6}\right]\ +\ \frac{\ell(\ell+1)}{2r^2}
	\label{Rydberg}
\end{equation}
where
\begin{equation}
	Z_\ell(r)\ =\ 1\ +\ (Z-1)e^{-a_1r}\ -\ r(a_3+a_4 r)e^{-a_2r}
	\label{ScreeningCharge}
\end{equation}
where   $\{a_1,a_2,a_3,a_4,r_c,a_c\}$ are parameters with dependence on the value of $\ell$ (angular momentum) and the atomic number $Z$. Potential (\ref{Rydberg}),  written in a.u, is useful to study the so-called Rydberg atoms. It describes a highly excited  valence electron through a radial potential, see Ref. \cite{Marinescu}. The spin-orbit interaction has been dropped \textcolor{black}{out} for simplicity.
For concreteness, we focus on the $p$-states associated to the valence electron of the Rubidium (Rb) atom, thus,  we take $Z=37$ and $\ell=1$. The values of the parameters in a.u. that characterize (\ref{Rydberg}) and (\ref{ScreeningCharge}) are the following\footnote{See Ref. \cite{Marinescu}.}
\begin{equation}
	\begin{split}
		a_c &= 9.0760\\
		a_1 &=4.440\,889\,78\\
		a_3 &= -16.795\,977\,70
	\end{split}
	\hskip1cm
	\begin{split}
		r_c &= 1.501\,951\,24\\
		a_2 &= 1.928\,288\,31\\
		a_4 &=-0.816\,333\,14		\ .
	\end{split}
	\label{parameters}
\end{equation}
To calculate eigenvalues and eigenfunctions simultaneously, we use the block of code shown below, see  \hyperref[B18]{\Block{18}}.
\begin{figure*}[h]
	\centering
	\includegraphics[scale=0.9]{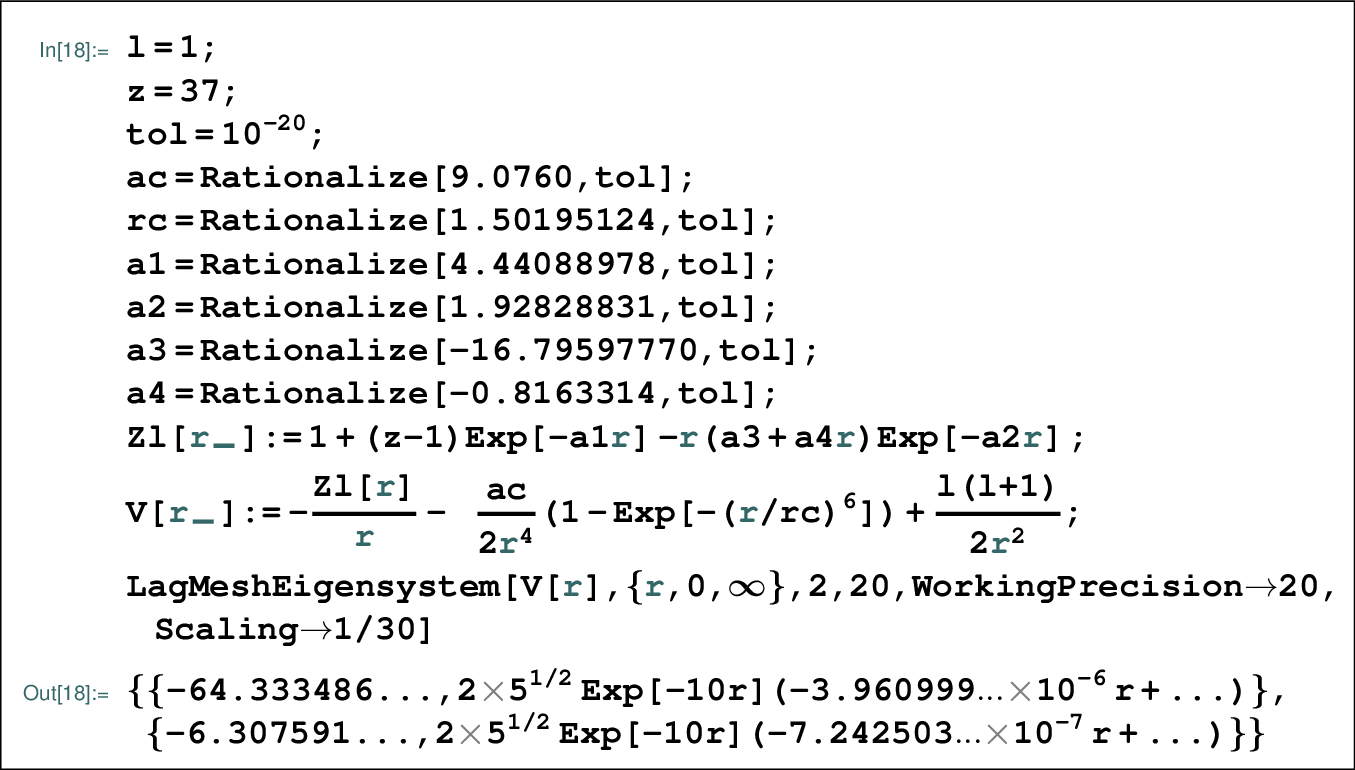}
	\label{B18}
\end{figure*}

A  Laguerre mesh with $N=20$ is more than sufficient to obtain the first two eigenvalues and eigenfunctions with accuracy consistent with the number of digits in the parameters (\ref{parameters}). In the previous calculation,  \mcom{Scaling} is crucial to obtain a good performance of the command \mcom{LagMeshEigensystem} as we now explain. First, note that the potential (\ref{Rydberg}) develops a deep well (depth $\sim-216$\,a.u.)  of small width $\sim 0.5$\,a.u. Therefore, the wavefunction is localized in a very small spatial region. Via \mcom{Scaling} we can move all mesh points to such region. This increases the  accuracy in results and reduces CPU times from minutes to a couple of seconds. 

Before concluding, there is a technical  issue that should be mentioned concerning potential (\ref{Rydberg}). Since exponential decaying  functions are involved,  the evaluation of potential matrix elements (\ref{VG}) can  lead to extremely   small numbers. This might lead to a loss of accuracy in calculations and to an increment in CPU times. This situation has  to be avoided as much as possible.  For the particular previous case,  \mcom{Scaling$\rar$1/30} solves the problem. 

\subsubsection{Possible Issues}
We give a brief review of common situations that the user may face. 

\paragraph{Loss of Accuracy.} In order to use arbitrary precision arithmetic offered by Mathematica$^\circledR$,  specified  by the option $\mtext{WorkingPrecision}$, it is mandatory to define all input parameters  $(\mtext{V[\textcolor{mygreen}{x}]}, \mtext{Dimension},\mtext{Mass},$ etc.$)$ with the same accuracy chosen for $\mtext{WorkingPrecision}$.  In general, defining all of them with absolute accuracy is more than enough. For example, instead of taking  \mcom{1.5},  use  \mcom{3/2}. Otherwise, accuracy in arithmetic will drop  to \mtext{MachinePrecision}.  Avoiding extremely small/large numbers can help  the performance as well. For example,   \mcom{Mass$\rar$9.1093837015$\times$10$^{\mcom{-31}}$} (kg), which is the mass of the electron, will lead to a poor performance of the package. Therefore, it is advisable to use the natural scales of the system. 

\begin{figure*}[h]
	\centering
	\includegraphics[scale=0.9]{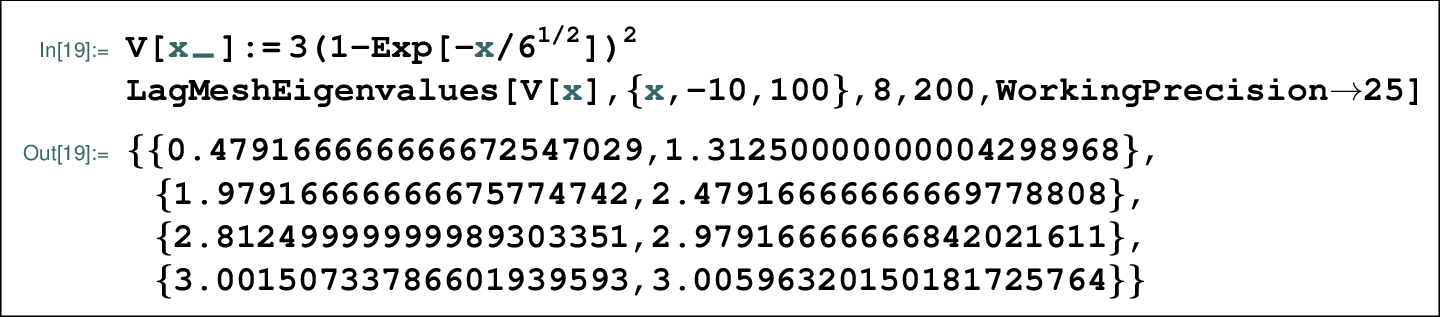}
	\label{B19}
\end{figure*}

Another situation that leads to a limited accuracy in the spectrum occurs when the wavefunctions are very extended in the domain. This is typical for weakly bound states defined on semi-infinite or infinite domains. In such situation, mesh points have to be adjusted to lie on the region where the wavefunction extends. Furthermore, when the potential $V(x)$ holds a finite number of bound states, the LMM usually predicts eigenvalues and eigenfunctions beyond the threshold. They correspond to approximate continuum states. To illustrate both situations, let us take the exactly solvable Morse potential:
\begin{equation}
	V(x)\ =\ 3\left(1 -e^{-\frac{1}{\sqrt{6}}x}\right)^2\ ,\qquad -\infty<x<\infty\ .
\end{equation}
This potential only holds six bound states, whose energies are given by
\begin{equation}
	E_n\ =\ \left(n+\frac{1}{2}\right)\ -\ \frac{1}{12}\left(n+\frac{1}{2}\right)^2\ ,\qquad n\ =\ 0,1,...,5\ .
	\label{Morse}
\end{equation}
Now, we calculate the spectrum using \hyperref[B19]{\Block{19}}.
\textcolor{black}{The domain specified  in \hyperref[B19]{\Block{19}} is suitable to study the fifth excited state, which is a weakly bound state whose wavefunction is considerably extended inside  $-3\leq x\leq 40$. We have intentionally requested eight eigenvalues to see explicitly those corresponding to (approximate) states in the continuum (the last two in \hyperref[B19]{\Blocko{19}}). Comparing numerical results with the exact spectrum (\ref{Morse}), the deviation is of order $10^{-12}$ or less. }

\paragraph{CPU times and RAM.} Computation time may become significant in different situations. Usually, either  large values for the option \mcom{Dimension} or \mcom{WorkingPrecision}  require a considerable CPU time. In particular, constructing accurate and large meshes via \mcom{BuildMesh} can take days of calculations. Choosing an appropriate value of \mcom{Scaling}, instead of increasing the size of the mesh, can lead not only to  an improvement of results, but also to a reduction of CPU times. When few eigenvalues or eigenfunctions are of interest, the option \mcom{Method}$\rar$\textcolor{gray}{\mcom{"Arnoldi"}} usually reduces CPU times drastically. Let us point out that  an increase in  \mcom{WorkingPrecision}  will certainly increase the arithmetic accuracy. However, the RAM needed to manipulate and store those  numbers during calculations will become larger. It is highly recommendable to keep  track  of RAM consume from the terminal.  

\paragraph{Limitations.}

\begin{figure*}[h]
	\centering
	\includegraphics[scale=0.9]{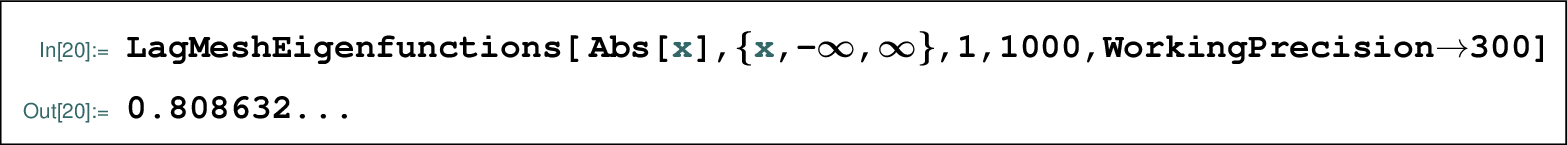}
	\label{B20}
\end{figure*}

 Any approximate method to solve the Schr\"odinger-type equations has its own \textit{domain of applicability}. The LMM is not the exception. One important limitation comes from the Gauss quadrature approximation. When the potential is not smooth, the Gauss quadrature fails and, as a result, matrix elements $V_{i,j}^{(G)}$  may be inaccurate. Certainly, this will lead to a poor accuracy in the  spectrum. A simple representative example of this situation occurs for the one-dimensional potential 
\begin{equation}
	V(x)\ =\ |x|\ ,\qquad-\infty < x<\infty\ ,
\end{equation}
which is not smooth at $x=0$. Using \mcom{LagMeshEigenvalues}, even with a large number of mesh points, eigenvalues are inaccurate, see  \hyperref[B20]{\Block{20}} and \hyperref[B20]{\Blocko{20}}. 

If we compare it with the \textit{exact} lowest eigenvalue $E_0\ =\ 0.808\,616\,...$,
we can immediately note the poor performance of the \mtext{LagrangeMesh} package. 

\section{Conclusions}
Through the years, many efforts have been put  into developing numerical methods to solve the time-independent one-dimensional Schr\"odinger  equation.  In this work, we focused on one of these methods to study bound states and approximate states in the continuum: the Lagrange Mesh method (LMM).
Specifically, the LMM is an approximate variational method simplified by a Gauss quadrature associated with a given mesh.  The method is based on three ingredients: Gauss quadrature approximation, Lagrange functions, and  secular equations.
The  non-existent ideal numerical method would be characterized by (i) short CPU times; (ii) high accuracy; (iii) applicable to any potential.
Under some  conditions, the LMM fulfills these three characteristics.

After presenting a comprehensive and detailed  discussion of the LMM, we introduced the \mcom{LagrangeMesh} Mathematica$^\circledR$ package. Once installed or loaded, it provides the user three  commands that  implement  the LMM numerically: \mcom{LagMeshEigenvalues}, \mcom{LagMeshEigenfunctions}, and \mcom{LagMeshEigensystem}.
The output of each one is the following: eigenvalues, eigenfunctions, and eigenvalues and eigenfunctions, respectively. All of them are delivered on screen in such a way that they are ready-to-use. 
Therefore, we have a powerful tool to find the quantum spectrum of an arbitrary potential defined on a given interval (which can be finite, semi-infinite, or infinite). Two complementary commands, \mcom{BuildMesh} and \mcom{AvailableMeshQ}, are  available to the user to  construct and check meshes and weights. All five  commands are user-friendly and none of them requires more than a couple of lines to specify the input.
The main properties of the commands  offered by the package are (i) efficiency; (ii) control  of  the accuracy in arithmetic manipulations and final results. These properties are controlled by three options \mcom{Dimension}, \mcom{Scaling}, and \mcom{WorkingPrecision}. 

Based on different relevant physical systems (\textit{worked examples}), we discussed  in detail the usage, scope, and limitations of the package. In particular, for some of those systems, we showed explicitly how benchmarks in eigenvalues are established in few lines of code in short CPU times. For example,  we established the most accurate results for the first energies of the  $\mathcal{PT}$-symmetric cubic potential. The collection of worked examples  presented in the text can be regarded as a user guide of the package. Thus, we hope that the \mcom{LagrangeMesh} package may serve as a tool  for educational and research purposes. In particular, it can be used to test other  highly accurate methods of different nature. 

Finally, the  LMM can be extended to tackle the time-independent S\"chrodinger equation with more than one degree of freedom/coordinate. For few-body systems, it has been shown several times that the method performs outstandingly, frequently leading to benchmark results, see Refs. \cite{Neutral,Helium}.  The LMM is extended by assuming that the wave function can be approximated by a linear combination of factorizable Lagrange functions with respect to  each coordinate. Thus, the mesh points are taken according to the range of each coordinate. Further details can be checked in Ref. \cite{Baye2015}. Concerning the \mcom{LagrangeMesh} package, it can certainly be  extended to tackle the multidimensional problem. In this case, the geometry of the system, usually  reflected in the choice of coordinates, plays a crucial role in the performance of the method. Covering most of the relevant systems of coordinates (Cartesian, polar, spherical, parabolic, etc.) is a challenge.  \textcolor{black}{However, it is worth mentioning that arbitrary precision arithmetic can be implemented in most of the programming languages used for scientific computing nowadays. In particular, an open-source but still user-friendly extension  of the  \mcom{LagrangeMesh}  working with arbitrary precision could be even more attractive and efficient for the scientific community.    }

\section*{Acknowledgments}
I am very grateful to H. Olivares-Pil\'on for the outstanding and comprehensive lectures about the LMM. I thank  A.V. Turbiner for the constant encouragement in developing  the package, many useful discussions, comments, and suggestions. I thank  S. C\'ardenas-L\'opez  and J.C. L\'opez-Vieyra for useful remarks and tests of the package.  I would like to thank K. Kropielnicka for valuable suggestions. This work was partially supported by the Simons Foundation Award No. 663281 granted to the Institute of Mathematics of the Polish Academy of Sciences for the years 2021-2023. Additional financial support was provided  by the SONATA BIS-10 grant no. 2019/34/E/ST1/00390.




\bibliographystyle{apsrev4-1}
\bibliography{references.bib}



	
	
	

\end{document}